\documentclass[letterpaper]{article} %
\usepackage[preprint]{aaai2027}
\usepackage[hyphens]{url}  %
\usepackage{graphicx} %
\usepackage{natbib}  %
\usepackage{caption} %
\usepackage{array}
\usepackage{booktabs}
\usepackage{amsmath}
\usepackage{float}
\newcommand{\bench}{\textsc{ComtradeBench}}
\newcommand{\audit}{\textsc{BenchAudit}}
\newcommand{\gtf}{ground-truth $F_1$}

\title{The Double Measurement Confound in Agent Benchmarks:\\
De-Scaffolding, Ground-Truth Scoring, and Reliability Beyond the Mean}
\author{
    Yonghong Zhang\textsuperscript{\rm 1}\corresponding,
    Shadi Motaali\textsuperscript{\rm 1},
    Vu Phong Dinh\textsuperscript{\rm 2},\\
    Avin Piroutiniya\textsuperscript{\rm 3},
    Jorge E. L\'opez de Vergara\textsuperscript{\rm 1},
    Luis de Pedro\textsuperscript{\rm 1},\\
    Ricardo Correia\textsuperscript{\rm 1},
    Isabel M. Parra\textsuperscript{\rm 1},
    Yong Xie\textsuperscript{\rm 4}\corresponding
}
\affiliations{
    \textsuperscript{\rm 1}Universidad Aut\'onoma de Madrid, Spain\\
    \textsuperscript{\rm 2}IMDEA Nanociencia, Spain\\
    \textsuperscript{\rm 3}Universidad Complutense de Madrid, Spain\\
    \textsuperscript{\rm 4}Spanish National Research Council (CSIC), Madrid, Spain\\
    yonghong.zhang@estudiante.uam.es, xieyong.nwpu@gmail.com
}

\begin{document}
\maketitle

\begin{abstract}

Agent benchmarks are increasingly used to compare large language models (LLMs) and guide deployment decisions, yet benchmark scores are meaningful only if they measure model capability rather than properties of the evaluation pipeline. We identify a \emph{double measurement confound}: execution-critical decisions are performed by a fixed scaffold instead of the model, while the scorer evaluates outputs using criteria that may not reflect task correctness. We unify these issues within a measurement-theoretic framework that characterizes when benchmark scores can be interpreted as evidence of model capability, and instantiate it with an audit-and-repair protocol that (i) transfers execution-critical decisions from the scaffold to the model, (ii) replaces shape-based evaluation with seeded ground-truth scoring, and (iii) reports reliability beyond the mean through worst-case and tail-risk metrics. Experiments on \bench{} show that the joint intervention transforms a nearly flat leaderboard into a reliability spectrum that distinguishes both average performance and robustness across seeds. Applying the audit to existing benchmarks further shows that scorer validity is
benchmark-specific, whereas scaffold ownership is an uncontrolled axis wherever we probed it. Our results suggest that benchmark scores should be interpreted together with their scaffolding level, scoring criterion, and reliability profile, providing a practical framework for more valid evaluation of LLM agents.

\end{abstract}

\section{Introduction}

\begin{figure*}[t]\centering
\includegraphics[width=0.72\textwidth]{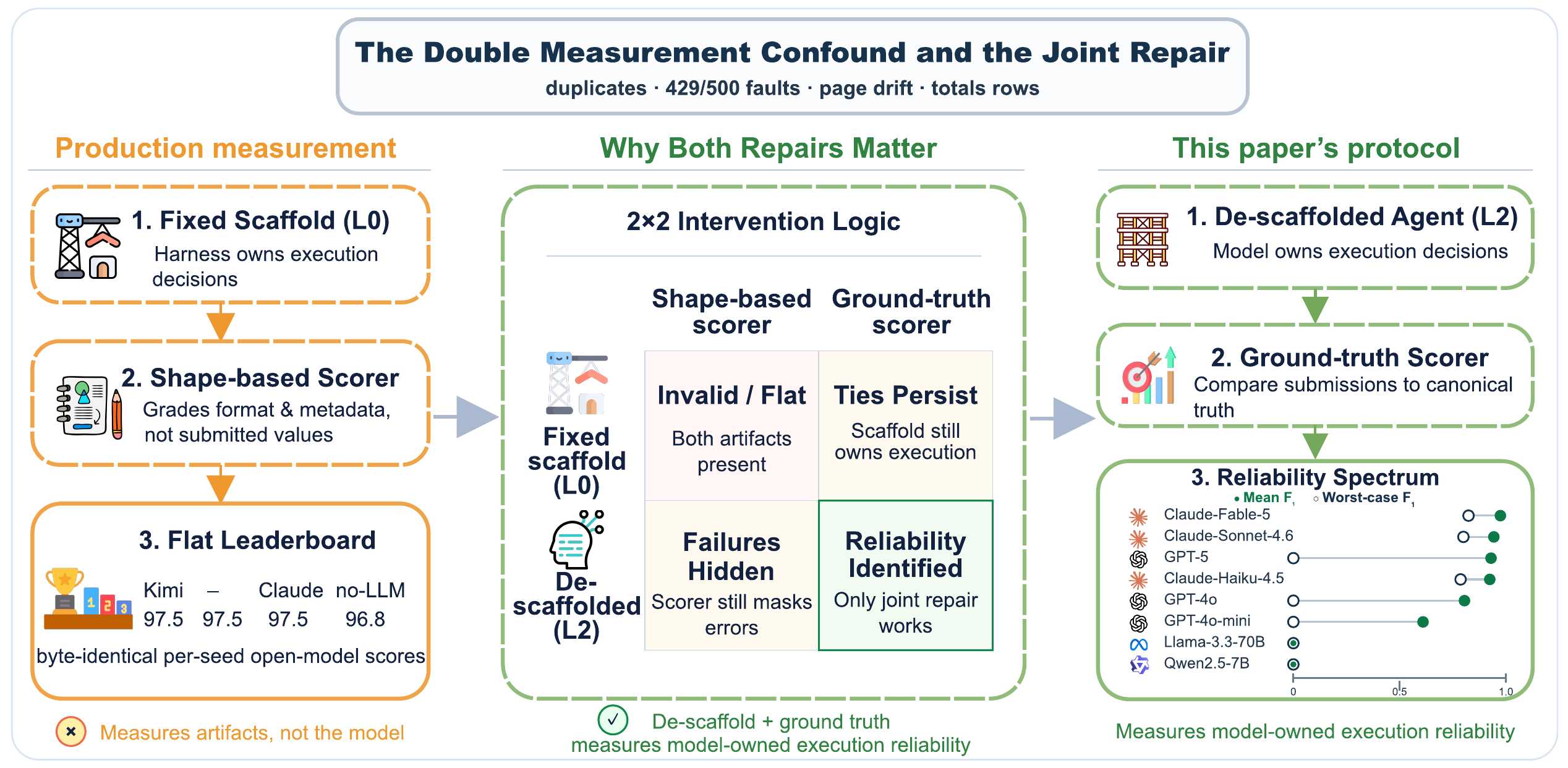}
\caption{The double measurement confound and the joint repair. \emph{Left}
(production): a fixed scaffold (L0) owns the execution-critical decisions and
a shape-based scorer grades format and metadata, so a no-LLM script ties the
frontier. \emph{Center}: why the two artifacts must be removed jointly
(Section~\ref{sec:joint}). \emph{Right} (our protocol): a de-scaffolded agent
(L2, Section~\ref{sec:scaffolding}) scored as \gtf{} (Section~\ref{sec:gtmetric})
against seeded ground truth, resolving Figure~\ref{fig:reliability}. Schematic;
every quoted number appears in the text.}%
\label{fig:framework}\label{fig:confound-paths}
\end{figure*}

Benchmarks for large language model (LLM) agents evaluate a complete
execution system that includes the model, tools, scaffold, environment, and
scorer. The reported score may therefore depend on the evaluation procedure as
well as the underlying model. Prior work has separately shown that harness
choices can change agent performance
\citep{yao2026harnessbench,kapoor2025hal}, automatic evaluators can disagree
with task correctness \citep{vaghasiya2026validity,xiao2023evaluating}, and
aggregate success rates can conceal unreliable behavior
\citep{yao2024tau,rabanser2026science}. What remains unclear is how these
measurement problems interact and when an agent benchmark can support claims
about model capability.

This issue is pronounced on a published benchmark for adversarial,
trade-data-shaped tool use. Two frontier models from independent labs, Kimi
and Claude, obtain identical mean scores of $97.5$, while a rule-based script
with no LLM scores $96.8$. Two unrelated open models, Qwen2.5-7B and
Llama-3.3-70B, also receive byte-identical per-seed scores. At this
configuration, the benchmark has little power to distinguish model capability
\citep{raji2021everything,bowman2021fix,liao2021learning,
reuel2024betterbench}.

We trace this result to a \emph{double measurement confound} involving the
scaffold and the scorer. The \textbf{scaffold} makes every execution-critical
decision, including retry, pagination, deduplication, and submission. The score
therefore reflects execution performed by the harness rather than the model.
The \textbf{scorer} evaluates output shape and self-reported metadata instead
of comparing the submitted values with ground truth
\citep{messick1995validity,jacobs2021measurement,xiao2023evaluating}.
Consequently, a fabricated record set receives the same score as the correct
one, $0.987$ in both cases. These artifacts also conceal one another. A valid
scorer still measures decisions made by the scaffold, while a model-controlled
execution remains mismeasured by an invalid scorer. We formalize this
interaction as a joint identification problem.

We transfer the execution-critical decisions to the model and score its
submission by \gtf{} (Section~\ref{sec:gtmetric}). This joint intervention
turns eight previously indistinguishable models into a graded spectrum, from
Claude-Fable-5 at $0.974$ to Llama-3.3-70B and Qwen2.5-7B at $0.000$
(Figure~\ref{fig:reliability}). A paired seven-model $2{\times}2$ intervention
shows that neither change is sufficient alone. Under the production scaffold,
payloads are byte-identical across all 21 model pairs, and both scorers tie
every model. The spectrum appears only when the model controls execution and
the submission is scored against ground truth
(Table~\ref{tab:joint}). The benchmark predates this audit: its scaffold, scorer, and leaderboard were fixed before the double-confound hypothesis, making this a retrospective self-audit rather than a benchmark built to exhibit the pathology.

The repaired evaluation also reveals reliability differences hidden by the
mean \citep{henderson2018deep,rabanser2026science}. GPT-4o ranks fifth by mean
but reaches a worst-case score of $0.00$, whereas the smaller
Claude-Haiku-4.5 never falls below $0.787$. Our pre-registered stress test
\citep{nosek2018preregistration} is confirmed under its designated scorer, but
the effect disappears when the same episodes are evaluated against ground
truth. 

In external audits, the official $\tau$-bench scorer is outcome-based, yet we find its four native scaffolds
shift the same model's reward by up to $0.267$ and change model rankings
\citep{yao2024tau}. The Berkeley Function-Calling Leaderboard (BFCL) instead offers an
abstract syntax tree (AST) track that scores a call against a function specification rather than
execution success \citep{patil2025bfcl}.

We make three contributions. First, we identify the \emph{joint} measurement
confound created by scaffold ownership and scorer validity, show that the two
artifacts mask each other so that repairing the scorer alone leaves every model tied while
de-scaffolding alone leaves the differences misgraded,
and state the condition under which an agent score identifies model behavior.
Second, we introduce \audit{}, an executable audit-and-repair protocol whose
deliverable is a per-benchmark \emph{validity card}, and demonstrate it on
three benchmarks; applied reflexively, it overturns the result of our own
pre-registered stress test. Third, we release \bench{} as a seeded generator
card together with the eight-model reliability spectrum, the paired
$2{\times}2$ intervention, and the $\tau$-bench and BFCL audits that
reproduce every number.

\section{Related Work}

A benchmark score is evidence about a model only when three things hold. The model, not the harness, makes the execution-critical
decisions. The scorer responds to whether the work is correct,
not only to how it looks. The report shows how scores vary
across runs, not only their mean. Prior work has largely developed each of
these separately. We review four bodies of work below, one per condition
plus documented scaffold sensitivity, and close with the non-stationary reinforcement-learning (RL) literature our apparatus inverts. Across the seven suites in our positioning
matrix (supplementary material) we find none that both reports seeded
worst-case reliability and treats the scaffolding level as a controlled
variable: $\tau$-bench's $\mathrm{pass}^k$ covers the first without seeded worst-case
estimation, and Harness-Bench's $6{\times}8$ factorial \citep{yao2026harnessbench}
covers the second over complete harnesses rather than a decomposed ownership
axis.

\paragraph{Tool-use and agent benchmarks.}
ToolBench \citep{qin2023toolllm}, BFCL \citep{yan2024berkeley,patil2025bfcl}, API-Bank \citep{li2023apibank}, AgentBench \citep{liu2023agentbench}, $\tau$-bench \citep{yao2024tau}, SWE-bench \citep{jimenez2024swebench}, and GAIA, WebArena, and OSWorld \citep{mialon2024gaia,zhou2024webarena,xie2024osworld} share two properties we build against. First, their adversity, where present, lives in the prompt or task specification; \bench{} places it in the environment's response dynamics, where it cannot be prompted away ($\tau$-bench is the closest relative, but its adversity is rule- and policy-level, not environment-level fault injection with within-episode dynamics). Second, they report means (or best-of-$k$ $\mathrm{pass}@k$) over few runs, and none treats the \emph{scaffolding level} as controlled. SWE-bench scores move substantially when the harness changes while the model is held fixed \citep{yang2024sweagent,xia2024agentless}; holistic leaderboards run the same models across several scaffolds \citep{kapoor2025hal}; and Harness-Bench \citep{yao2026harnessbench} runs a $6{\times}8$ harness--model factorial while explicitly \emph{not} decomposing individual harness mechanisms. We formalize that missing decomposition into a controlled \emph{ownership} axis, with seeded repeats that support worst-case reporting.

\paragraph{Reliability and worst-case evaluation.}
We import a tail-risk vocabulary under-applied to LLM-agent evaluation: CVaR from finance \citep{rockafellar2000optimization}, carried into risk-sensitive and distributional RL \citep{chow2015risk,bellemare2017distributional}; worst-group accuracy from distributionally robust learning \citep{sagawa2020distributionally}; and multi-seed variance and confidence-interval reporting \citep{agarwal2021deep}. Two points sharpen our position: the worst case \emph{reorders} the mean ranking up to the frontier (a cross-model \emph{re-discriminator}, not only a within-model tool), and where worst-group robustness conditions on known subpopulations, we condition on seeds of a seeded adversary, making the worst case estimable rather than anecdotal.

\paragraph{Benchmark validity and automatic scorers.}
A separate line asks whether a benchmark score means what it is read to mean:
construct and criterion validity in measurement theory
\citep{messick1995validity,campbell1959convergent}, its transfer to machine
learning \citep{jacobs2021measurement,hutchinson2022evaluation}, benchmark-level
critiques \citep{raji2021everything,bowman2021fix,liao2021learning,
reuel2024betterbench}, and analyses of automatic evaluation metrics and
prompt-dependent scores \citep{xiao2023evaluating,mizrahi2024state}. This work
establishes that scores can fail to track the construct; it does not ask
\emph{who executed} the task being scored. We treat scorer validity and
scaffold ownership as a single identification problem, and make the resulting
condition auditable rather than argued.

\paragraph{LLM-agent variance and scaffolding sensitivity.}
LLM agents are documented as high-variance across seeds, prompts, and harnesses \citep{miller2024adding,kapoor2024ai,sclar2024quantifying}, and the scaffold can shift measured success as much as the model \citep{yang2024sweagent,xia2024agentless}. Prior work \emph{observes} the variance as a confound; we make it the \emph{object of measurement} via a controlled scaffolding spectrum, trace a large share of the apparent variance to the scaffold's hidden execution work (Section~\ref{sec:confound}), and show the diagnosis is \emph{correctable}.

\paragraph{Non-stationary and continual RL.}
Concept drift \citep{gama2014survey}, non-stationary bandits
\citep{besbes2014stochastic}, continual RL \citep{khetarpal2022towards}, and
hidden-mode POMDPs \citep{choi2000hidden} study environments that change while
a learner adapts. Our apparatus inverts the object: the agent is \emph{frozen}
and the question is whether a fixed policy executes reliably under
within-episode adversity, with escalation driven by agent success rather than
elapsed time \citep{lee2020adaptive}.

\section{The Double Measurement Confound}\label{sec:confound}

\bench{} is a seeded benchmark of adversarial, paginated data extraction,
themed on public trade-statistics records: given a query and an operational
budget, an agent must fetch all pages of a mock data API, deduplicate
records, drop summary rows, retry transient faults, and submit a cleaned
record set with metadata and a run log. The adversity is \emph{in the
environment} (faults, duplicates, and distractor rows are injected by the
mock service, not described in the prompt), so it cannot be reasoned around.
Scope: \bench{} is stylized adversarial extract-transform-load
 work themed on trade data; it
tests execution reliability, not economic reasoning. All data
is a deterministic function of \texttt{(task\_id, seed)}, so ground truth
is regenerable from the seed (Section~\ref{sec:protocol}).
Ten tasks isolate fault families, with a fault-free variant
(\texttt{T9\_clean\_ab}) as the substrate for the A/B design of
Section~\ref{sec:experiments}; per-task parameters are in the supplementary
material. The agent interacts through three model context protocol (MCP) tools, constant across
all scaffolding levels: \texttt{get\_task\_info()},
\texttt{fetch\_page(page, page\_size)}, and
\texttt{submit\_results(data, metadata, run\_log)}. The tools are fixed;
what varies is the \emph{wrapper} around them: the first artifact.

\subsection{Artifact 1: The Scaffold Does the Work}

In the production scaffold, the agent loop wraps the tool calls in fixed
control logic: it auto-retries transient faults with backoff, deduplicates
fetched rows by primary key, drops totals rows, loops pagination to
completion, and assembles and submits the final payload. These are the
execution-critical decisions the adversarial API is designed to stress.
When the scaffold owns them, the LLM only fills templated slots (which query to run, when to stop paging).

The resulting invariance is not approximate. On the published leaderboard,
Kimi and Claude post numerically identical mean scores (97.5 each), and a
rule-based baseline with no LLM at all scores 96.8, within roughly 0.7
points of the frontier agents. At the seed level the
invariance is exact: under the production agent, Qwen2.5-7B and
Llama-3.3-70B produce \emph{byte-identical per-seed reward vectors} on the
multi-page task.

\subsection{Artifact 2: The Judge Scores Shape}

The benchmark's deployed scorer (\texttt{judge.py}) is a deterministic,
100-point rubric over six dimensions: correctness (30), completeness (15),
robustness (15), efficiency (15), data quality (15), and observability (10),
with two governance gates.
This is a criterion-validity question: \emph{the judge scores
output shape and self-reported metadata, never the submitted values against
ground truth}. Correctness grades the submitted-line \emph{count} against the
submitter's own \texttt{metadata.row\_count} (never whether the ids are the
true ids), robustness rewards \emph{logging} a retry rather than retrieving
records, and no dimension compares the submission to the seeded ground truth
(full per-dimension walkthrough, gates included, in the supplementary
material). A
probe in the supplementary material passes three canned submissions through the
released \texttt{judge.py} and reports each six-dimension breakdown: an
empty \texttt{data.jsonl} with well-formed metadata and a disciplined run
log scores $0.648$, and a fabricated record set with invented ids and
self-consistent metadata receives the \emph{identical} judge total as the
correct ground-truth record set ($0.987$ each, at \gtf{} $0.0$ versus
$1.0$). Metadata-shape and
log-discipline credit floor a contentless submission at one half or above,
while the rubric compresses real differences at the top. The released pipeline's probe battery (Section~\ref{sec:protocol}) adds the converse test: the \emph{same} correct records paired with a contradictory self-report drop by $0.398$, while fabricating the data behind a well-formed report costs $0.000$. The judge grades the report, not the data.

\subsection{Why the Artifacts Must Be Removed Jointly}
\label{sec:joint}

The two artifacts mask each other
(Figure~\ref{fig:confound-paths}). Fix the scorer alone and it faithfully
reports the scaffold's near-perfect execution: the model still owns no
decision the adversary targets, so the model-invariant ties persist. Remove
the scaffold alone and the shape-based judge hides the model's real failures
on the way out, flooring the empty submission, ceilinging the perfect one,
and compressing the genuine spread into a band that a leaderboard reads as noise.
Only joint removal recovers a measurement of the model: return retry,
deduplication, totals filtering, and the submit decision to the model, and
score the submitted unique records against the seeded ground-truth record
set, with duplicates and totals rows counted as errors; the seeded
generator makes this scorer free and deterministic per
\texttt{(task, seed)}. Section~\ref{sec:protocol} specifies the resulting protocol. We keep the six-dimensional judge in the released
benchmark as its default scorer and as an object of study: the
judge--ground-truth contrast is itself the evidence for the
measurement-validity claim. Section~\ref{sec:experiments} tests this as a $2{\times}2$ intervention
(Table~\ref{tab:joint}).
\section{Measurement Protocol}
\label{sec:protocol}

We now specify the protocol; the supplementary material diagrams the full
pipeline alongside the measured L0/L1/L2 spectrum, and
Figure~\ref{fig:confound-paths} contrasts its scoring path with the
production path.

\subsection{A Measurement Model}
\label{sec:model}

Write a benchmark score as $S(m, c, j)$: the reward model $m$ receives when
run under scaffold $c$ (a harness owning some subset of the
execution-critical decisions of Section~\ref{sec:confound}) and graded by
scorer $j$. A leaderboard reports $S(\cdot, c_0, j_0)$ at one fixed
configuration and interprets it as a property of $m$ alone \citep{mizrahi2024state}. Call the
configuration \emph{model-invariant} if $S(m_1, c_0, j_0) = S(m_2, c_0,
j_0)$ for all models under test, and call $j$ \emph{criterion-valid} if $S$
is strictly increasing in the ground-truth quality of the submitted work
product.

\textbf{Proposition 1 (non-identifiability).} If (i) the scaffold owns every
execution-critical decision, so that no scored quantity depends on a model
choice beyond templated slots, and (ii) the scorer grades only submission
shape and self-reported metadata, then $S(m, c_0, j_0)$ is invariant in $m$
whenever two models fill the template equivalently, and no function of the
leaderboard identifies a property of the model. Both hypotheses and the
conclusion hold at the production configuration (Section~\ref{sec:confound}):
the invariance is realized byte-for-byte across two unrelated open models. The
protocol below is the identification strategy: move $c$ to L2 so the decisions
depend on $m$, and $j$ to ground truth so the score depends on the decisions;
dependence on $m$ then reappears (Figure~\ref{fig:reliability}).

\textbf{Proposition 2 (identification).} Conversely, if every
execution-critical decision is model-owned and the scorer is
criterion-valid, then the score is a deterministic function of the model's
decisions in the seeded environment, so any difference in $S$ between two
models at matched seeds reflects a difference in model behavior, and
seed-level statistics (worst case, CVaR) estimate the model's execution
reliability. What behavior is measured remains level-dependent: at L2 the
identified quantity includes the re-emission demand, the mechanical-floor
caveat of Section~\ref{sec:scaffolding}.

The propositions yield a minimal condition a benchmark can be audited
against. Write $D_{\mathrm{claimed}}$ for the decisions the benchmark's
construct claims to measure and $D_{\mathrm{model}}$ for the decisions its
harness actually leaves to the model; the capability claim is interpretable
only if $D_{\mathrm{claimed}} \subseteq D_{\mathrm{model}}$, and when the
inclusion fails there are exactly two repairs: return the decisions to the
model, or relabel the score as system-level performance \citep{hutchinson2022evaluation}. Ownership is necessary, not sufficient: a model-owned channel must also be \emph{non-degenerate}, with $\operatorname{Var}(A) > 0$ over the chosen
actions and $E[Y \mid A{=}a_1] \neq E[Y \mid A{=}a_2]$ for some action
pair. Our L1 instrument fails exactly this check (SKIP chosen 0 of 120
episodes; Section~\ref{sec:escalation}), and no sample size repairs it:
the effect is non-identified, not merely low-powered \citep{card2020little}.
Rule-policy calibration (always-retry, always-skip, quota-aware) is the
pre-run check.

\subsection{The Scaffolding Spectrum}
\label{sec:scaffolding}

A benchmark evaluates a model \emph{inside a harness}, and the harness makes
the execution-critical decisions of Section~\ref{sec:confound}. We call this
bundle the \emph{scaffold} and treat the share of it owned by harness versus
model as a controlled variable with three reference levels.

\textbf{L0: production auto-everything.} The harness performs every
execution-critical step; the model fills templated slots and owns no decision
the adversary targets. L0 is a common shipping configuration; it
maximizes mean score and minimizes model signal, producing the ties of
Section~\ref{sec:confound}.

\textbf{L1: half-scaffold.} The harness handles the mechanical bottleneck
(pagination, writing the final record file); the model owns the one
adversary-relevant decision: on a fault, retry or skip.

\textbf{L2: full slim.} The model owns everything, including re-emitting
the cleaned record set and calling \texttt{submit\_results}. L2 is where the
populated reliability spectrum of Figure~\ref{fig:reliability} appears.

L0 hides all model differences; L2 exposes the spectrum but, for the weakest
models, folds ``decided wrongly under adversity'' and ``failed the mechanical
re-emission'' into one score; L1 separates them, and shows the isolated
decision to be degenerate here (Section~\ref{sec:experiments}): L0 and L2
are the informative extremes, L1 the diagnostic explaining the floored
zeros. A number quoted without its scaffolding level is not
interpretable.

\subsection{The Ground-Truth Metric}
\label{sec:gtmetric}

The shipped judge does not support a reliability claim
(Section~\ref{sec:confound}). For all headline numbers we score against the seeded canonical record set.
The generator emits each record with a deterministic id
(\texttt{\{task\}-\{i:06d\}}), so the true unique-record set for any (task,
seed) is known exactly. With $S$ the submitted records, $G$ the ground-truth
set, and $U$ the distinct ground-truth records in $S$, $R = |U|/|G|$,
$P = |U|/|S|$, and $F_1 = 2RP/(R+P)$: bounded in $[0,1]$ and deterministic
per (task, seed). Unlike the judge, it
drops when the data is wrong: omitted real rows cut recall; duplicates, totals
rows, and fabricated rows inflate $|S|$ but not $|U|$, cutting precision.

\bench{} is issued as a \emph{generator card}, a datasheet for a seeded
procedural generator rather than a static corpus
\citep{gebru2021datasheets}.
Because a run reproduces from only the task id and the seed, exact test-item
memorization is less likely, though generator- and template-level contamination
remain possible \citep{zhu2024dyval,oren2024proving,golchin2024timetravel,jacovi2023stop}.

\subsection{Reliability Metrics}
\label{sec:metrics}

The quantity that pages a human is not the mean but the worst case across
the adversary's seeds. Each run is one (level, seed) $\to$ terminal reward,
where a level is an opaque difficulty knob. We report:
\emph{worst-case} (minimum over tested seeds, an estimate of the deployment floor);
\emph{CVaR@$\alpha$} (mean of the worst $\alpha$-fraction, $\alpha=0.2$
default) \citep{rockafellar2000optimization}; %
\emph{Reliability@$\tau$} (fraction of runs clearing threshold $\tau$;
$\tau=0.9$ default, also reported at $0.5$), akin in spirit to $\mathrm{pass}^k$
\citep{yao2024tau}; %
a \emph{seeded bootstrap 95\% CI} for the mean (2000 iterations,
bit-reproducible) \citep{efron1979bootstrap}; %
and, for paired A/B comparisons, the within-pair \emph{Cliff's $\delta$}
(dominance $[\#(\Delta_i{>}0)-\#(\Delta_i{<}0)]/n$, the effect-size
companion of the sign-flip test) with the same seeded bootstrap CI
\citep{cliff1993dominance}. %
Across difficulty levels we summarize the degradation curve and the
\emph{reliability gap}: the static mean minus the hardest-level worst-case,
the operative column of Section~\ref{sec:experiments}.

\subsection{The Non-Stationarity Engine}
\label{sec:engine}

A domain-agnostic engine generates the stress reproducibly and lets us test
whether its \emph{shape} matters, with per-step randomness seeded by
\texttt{sha256(seed:step)}: non-stationary does not mean non-reproducible. It
supplies the matched-mean escalating-versus-constant A/B design we pre-register
(Section~\ref{sec:experiments}); its architecture and scales are in the
supplementary material.

\textbf{The protocol as a reproducible audit pipeline.} The full protocol is
implemented as \audit: given a benchmark's harness, scorer, and seeded
tasks, it emits the scaffold-ownership map, the forged-submission scorer probe
(correct, empty, fabricated, duplicate-laden, contradictory-metadata), the
$2{\times}2$ joint intervention (Table~\ref{tab:joint}), de-scaffolded
reliability statistics, and a per-benchmark \emph{validity card} whose cells are
pass, fail, or \emph{not probed}. Agents play no role in scoring; every reported
number originates in the deterministic, seeded core, and a self-audit against
\bench{} reproduces each finding as an executable regression check. Three
optional, human-confirmed LLM-assisted layers \emph{propose} rather than score
(scaffold discovery, harness synthesis, probe design), gated by a number guard
that rejects any narrated figure absent from the audit JSON: an auditor that is
itself auditable.
\section{Experiments}
\label{sec:experiments}

We report three results on \bench{}, then audit two external benchmarks and
summarize every target as a validity card. All headline numbers
use the ground-truth scorer of Section~\ref{sec:gtmetric}. E1 runs all eight
models at the full de-scaffold level L2 on the five execution tasks of
Figure~\ref{fig:reliability}, 10 seeds per (model, task); the escalation
experiments use the L1 half-scaffold with a ground-truth coverage reward,
$n = 20$ seeds per arm, paired by seed. Every interval is a seeded bootstrap
and reproduces bit-for-bit.

\subsection{The De-Scaffolded Reliability Spectrum}

\begin{figure*}[t]\centering
\includegraphics[width=.85\textwidth]{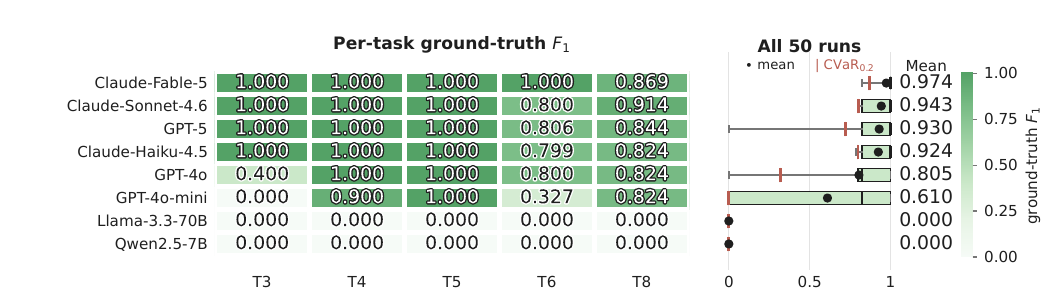}
\caption{E1: the de-scaffolded (L2) reliability spectrum. \emph{Left block}: per-task
\gtf{} (mean over 10 seeds) on the five execution tasks (T3 dedup, T4 429,
T5 500, T6 drift, T8 mixed). \emph{Right block}: all $n{=}50$ runs per model
--- box $=$ interquartile range, whiskers $=$ min--max, $\bullet$ $=$ mean
(also the numeric column), tick $=$ CVaR at $\alpha{=}0.2$; rows sorted by
mean. The lower whisker end is the worst-case, and it reorders the ranking
the mean implies: GPT-5 and GPT-4o post high means ($0.930$, $0.805$) yet
whiskers reaching $0.00$, while the far smaller Claude-Haiku-4.5 never drops
below $0.787$. Per-task dispersion, the reliability gap, and the excluded T7
diagnostic are in the supplementary material.}
\label{fig:reliability}
\end{figure*}

Removing both artifacts of Section~\ref{sec:confound} (the LLM owns retry,
dedup, totals filtering, re-emission, and submission, scored as \gtf{}) opens
the flat leaderboard into a populated spectrum (Figure~\ref{fig:reliability}).

The spectrum spans its full range (Figure~\ref{fig:reliability}).
Claude-Fable-5 leads on \emph{both} mean ($0.974$) and worst case ($0.824$),
clearing every single-fault task and leaving discriminative headroom at the
top; a \emph{previous-frontier tier} (Claude-Sonnet-4.6, GPT-5, and the small
Claude-Haiku-4.5) clusters at mean $0.92$--$0.94$ but is no longer uniform in
reliability. A \emph{high-variance mid tier} (GPT-4o, GPT-4o-mini) pairs
strong means with a worst case of $0.00$, and a \emph{floored tier}
(Llama-3.3-70B, Qwen2.5-7B) scores $0.000$ throughout. The floored zeros are a
marshaling failure, not a decision-quality result: Llama fetches rows
($\approx 0.83$ coverage at L1) yet submits an empty payload at L2, and Qwen
never submits a valid one, which is why the scaffolding level travels with
every number.

The tail statistics of Section~\ref{sec:metrics} separate models the mean
does not. GPT-5 and Claude-Haiku-4.5 differ by $0.006$ in mean \gtf{}
($0.930$ versus $0.924$) but by $0.079$ in CVaR@$0.2$ ($0.720$ versus
$0.799$); at $\tau = 0.9$, Reliability@$\tau$ is $0.720$ for GPT-5,
$0.600$ for Claude-Haiku-4.5, and $0.480$ for GPT-4o. A mean-only report
would call the first two equivalent.

The headline is that \emph{worst-case reliability reorders models the mean
rates as capable}. GPT-4o ranks fifth by mean yet floors at worst case, through
genuine bimodality: on dedup (T3) its per-seed $F_1$ is
$[0,0,1,0,0,0,0,1,1,1]$, and a forensic replication ($n{=}30$) closes the
mechanism, the outcome fully determined by whether the model ever varies an
off-by-one \texttt{fetch\_page(page=0)} call ($14/14$ of solves, $0/16$ of
failures): identifiable error-recovery behavior, not seed noise. The reorder
reaches the frontier: GPT-5 is bimodal on mixed faults, handing the suite's
largest gap ($0.930$) to a model the mean ranks third, while the small
Claude-Haiku never drops below $0.787$. Reliability tracks neither raw size
nor mean score (full transcript taxonomy in the supplementary material).

\subsection{Scaffolding Level as a Hidden Axis}

The same environment, measured at the three levels of
Section~\ref{sec:scaffolding}, returns three different verdicts. At L0
(production auto-everything) all model differences vanish into the ties of
Section~\ref{sec:confound}: the score measures the scaffold. At L2 the
spectrum of Figure~\ref{fig:reliability} appears, but with a mechanical floor:
part of the floored-tier zero is the marshaling chore, not adversity. L1,
which hands the model only the single adversary-relevant decision (on a fault,
retry or skip), shows that even this decision is degenerate here: across 120
episodes, SKIP is chosen
\emph{zero} times, and coverage settles near $0.83$, model-invariant, capped
by duplicate-overwrite in the mock rather than by any model choice;
the always-retry substrate of the pre-registered null below. The
scaffolding level thus changes not just the magnitude but the
\emph{qualitative content} of what is measured.

The audit pipeline runs the axis as a $2{\times}2$ \emph{joint
intervention}: one batch (\texttt{T3\_duplicates}, five seeds per cell,
seven of the eight models across three providers; Qwen's provider was
unavailable), every episode scored by both the shipped judge and ground
truth, the harness swapped between L0 and L2 with nothing else changed.
Table~\ref{tab:joint} reports the cell means (drawn as a $2{\times}2$ panel in
the supplementary material). At L0 the submitted payloads are SHA-256-identical
across \emph{all 21 model pairs} on every seed, reasoning models included, so
both scorers tie all seven: fixing the scorer alone reveals nothing. De-scaffolding
alone still misgrades what it separates: among the models that submit, the judge
compresses a $1.000$ ground-truth spread to $0.425$, scoring Llama's all-wrong
submissions above zero. Only the joint cell recovers the spectrum ($0.000$ to
$1.000$), so the masking argument of Section~\ref{sec:joint} is measured, not
argued. Llama is served by a different provider here than in
Figure~\ref{fig:reliability} (BF16 versus FP8, disclosed) and floors regardless:
the L2 floor is not a serving-stack artifact.

\begin{table}[t]
\centering
\small
\setlength{\tabcolsep}{4pt}
\begin{tabular}{@{}lcccc@{}}
\toprule
 & \multicolumn{2}{c}{L0 (production)} & \multicolumn{2}{c}{L2 (slim)} \\
\cmidrule(lr){2-3} \cmidrule(l){4-5}
Model & judge & gt $F_1$ & judge & gt $F_1$ \\
\midrule
Claude-Fable-5   & 0.987 & 1.000 & 0.902 & 1.000 \\
Claude-Sonnet-4.6 & 0.987 & 1.000 & 0.948 & 1.000 \\
GPT-5            & 0.987 & 1.000 & 0.768 & 1.000 \\
Claude-Haiku-4.5 & 0.987 & 1.000 & 0.781 & 1.000 \\
GPT-4o           & 0.987 & 1.000 & 0.614 & 0.400 \\
GPT-4o-mini      & 0.987 & 1.000 & 0.000 & 0.000 \\
Llama-3.3-70B    & 0.987 & 1.000 & 0.523 & 0.000 \\
\midrule
Model spread     & 0.000 & 0.000 & 0.948 & 1.000 \\
\bottomrule
\end{tabular}
\caption{The $2{\times}2$ joint intervention: harness swap $\times$
scorer swap on shared episodes (\texttt{T3\_duplicates}, $n{=}5$ seeds
per cell, one batch, seven models over three providers; each episode
double-scored; model order as Figure~\ref{fig:reliability}). L0
submissions are sha-256-identical across all 21 model pairs on all five
seeds. GPT-4o-mini's L2 zeros are non-submission (both scorers agree);
among submitting models the judge compresses the $1.000$ ground-truth
spread to $0.425$. Models separate fully only when both artifacts are
removed (rightmost column).}
\label{tab:joint}
\end{table}

\subsection{The Confound Inside the Confirmatory Experiment: a
Pre-Registered Stress Test}
\label{sec:escalation}

This experiment closes the causal chain: the scorer half of the confound
resurfaces inside our own pre-registered test. We pre-registered the
hypothesis that escalating, progress-driven stress degrades agents more than
matched-mean constant stress (\texttt{PREREGISTRATION.md}, frozen 2026-06-17)
\citep{nosek2018preregistration}, and report it as scorer-confound evidence,
not a non-stationarity finding: it is confirmed under the scorer the
pre-registration named and vanishes under ground truth. The B2 follow-ups are
\emph{exploratory, not pre-registered}: the enforced-quota variant the
pre-registration explicitly deferred.

\textbf{B1: a null, with a disclosed instrument deviation.} On
\texttt{T9\_clean\_ab} the matched-mean design is null for all three models
($\Delta = +0.000$; Cliff's $\delta = +0.05$, 95\% CI $[-0.30,+0.40]$), and
SKIP, the only channel a model could move its score
through, fired in 0 of 120 episodes; the run used the L1 coverage instrument,
blind to the pre-registered dedup/totals channel (disclosed per the deviation
rule; powered against large effects only, see Limitations). The
pre-registration-faithful rerun (slim agent, terminal judge, $n{=}20$/arm) is
null for Llama and Claude but gives GPT-5 $\Delta = +0.011$, CI
$[+0.002, +0.023]$, $p = .034$, $\delta = +0.25\,[-0.15,+0.65]$: a
\emph{confirmation} under the
pre-registration's ``$\geq 1$ frontier reasoning model'' outcome. We do not
dismiss it as noise. The scorer explains it: on the identical episodes GPT-5's
\gtf{} and coverage deltas are $\approx 0$ and the submitted data is equally
correct across arms, so the confirmation lives \emph{entirely} in judge
dimensions ground truth does not measure: the scorer-dependence this paper
diagnoses, surfacing inside our own confirmatory experiment
(\texttt{results\_b1\_slim.json}).

\textbf{B2: an exploratory follow-up.} B2 enforces a
request quota with a one-shot mid-episode squeeze only the escalating arm
crosses; the penalty turns positive on all three models (Llama and GPT-5
survive Holm correction, Claude does not) and revives the SKIP channel
($0/120 \to 38/120$). But a no-LLM \emph{always-retry} policy reproduces the
flip at $\Delta = +0.096$, larger than every model's penalty: resource
arithmetic that model decisions only attenuate. What the benchmark measures
here is \emph{decision calibration, not capability}. Effect sizes, quota and
squeeze parameters, rule-agent calibration, the zero-slack invariance
argument, and the dose-response curves are in the supplementary material.

\subsection{External Validity: Scorer Probed, Scaffold Measured}
\label{sec:external}

Is the scorer half general? We probed $\tau$-bench's official scorer
\citep{yao2024tau} with zero model calls, replaying released trajectories
through its released reward function and perturbing only the semantics: a
verbatim passing trajectory scores $1.0$, while a corrupted database write
and a wrong answer, both shape-preserving, score $0.0$. The scorer is
outcome-based (it hashes the final database state), so the shape-scorer
artifact is benchmark-specific (method and a trivial-agent finding in the
supplementary material). \looseness=-1
The \emph{scaffold} half, however, is present there as an
uncontrolled axis, and we \emph{measured} it: three models from two
providers (Claude-Sonnet-4.6, Claude-Haiku-4.5, GPT-4o) across all four
scaffolds the harness ships (tool-calling, few-shot, act, react; the LLM
user simulator is Claude-Haiku-4.5 throughout, itself one of the
evaluated agents, which we disclose). Changing nothing
but the scaffold flag moves the same model's reward by up to $0.267$ and
reorders the models (GPT-4o ties Sonnet under act yet finishes last under
react), so a leaderboard that fixes any single scaffold can favor one model
over another \citep{alzahrani2024benchmarks}. We report the slice as estimation
only ($n{=}15$/cell, sign tests n.s., single-trial noise of the same order as
the spread; full matrix in the supplementary material). A second probe, on BFCL's AST
scorer \citep{patil2025bfcl} (400 entries, five canned submissions each,
offline), finds a third profile: the correct call passes in all 400 while
value corruption, fabrication, wrong names, and empty arguments all fail ($0$
of $2{,}000$ cells deviate), a \emph{specification}-based scorer certifying
spec validity rather than execution success. 

Across the three benchmarks the
scorer half is benchmark-specific, the scaffold half is uncontrolled wherever we probed it. Read as a two-factor design (four shipped scaffolds $\times$ three models,
15 tasks per cell), the same slice attributes $\eta^2 = 0.0293$ to the
scaffold main effect against $0.0080$ for the model. Both shares are small
against the task-level residual, and the design is single-trial, so we
read only their ratio and only as an ordering: on this benchmark's own
grid the measurement method accounts for more of the variance
than the property being measured. At \bench{}'s production configuration
the corresponding model share is identically zero, the per-seed vectors
being byte-identical.

\subsection{The Validity Card}
\label{sec:card}

The audit's deliverable is one card per benchmark; a \emph{not probed} item is
never read as passing. Two audits gate the card's \emph{licensed claim}: whether
the scorer is criterion-sensitive (it separates correct from degraded work
product) and whether the model owns the execution-critical decisions the
construct claims (Section~\ref{sec:model}). The strongest reading each
configuration permits follows a fixed lattice: model capability only when both
pass and the run replays; an outcome scorer over an uncontrolled scaffold, only
model--scaffold system performance ($\tau$-bench); a shape scorer, only
self-report compliance (\bench{} as shipped, repairable at L2); a specification
scorer with scaffold not probed, only specification compliance (BFCL). The full
per-item card is in the supplementary material
(\texttt{audits/validity\_cards.md}).

\section{Limitations}

\paragraph{B2 and B2-slack are exploratory.} The pre-registered result is the B1 null; B2 and its slack ablation are the deferred enforced-quota variant ($n = 20$/arm, three models, one task family), offered for confirmatory follow-up. The zero-slack flip is resource coupling, not decision failure: once the quota binds, retry and skip are outcome-equivalent, so it cannot indict any model's decisions, and we ran one slack level.

\paragraph{Single domain.} \bench{} is stylized trade-data-shaped extraction work from a seeded generator, and only \texttt{ComtradeAdapter} realizes the full ownership
intervention (external targets are scorer-audited): cross-domain transfer is a design property, not yet empirical.

\paragraph{A small suite with a measured scale bound.} The spectrum rests on five tasks $\times$ 10 seeds, two ceiling-saturated for the frontier. A sixth task (T7) is excluded by a solvability audit: its 750-row re-emission demand exceeds any single-submission budget ($F_1$ cap $0.26$/$0.46$) and floors every model run on it, where L2 scoring stops measuring decisions and starts measuring output budgets (supplementary material).

\paragraph{Scope and provenance.} B1 is powered only against large effects (power $0.998$ at $n = 20$), so smaller escalation effects are not excluded; per-model scores are point-in-time on pinned API endpoints.
\section{Conclusion and Future Work}

\looseness=-1
When the scaffold owns the execution-critical decisions and the scorer grades
shape, a benchmark score is not identified in the model: seven models submit
byte-identical payloads, and a fabricated record set scores as high as the
correct one. The joint repair recovers a spectrum in which the worst case
reorders the mean, and the same confound changes the conclusion of our own
pre-registered experiment. Four practices follow. \emph{Measure the model, not
the scaffold}; \emph{score against ground truth, not shape}; \emph{report
worst-case, CVaR, and Reliability@$\tau$ as first-class outputs}; and
\emph{treat the scaffolding level as a controlled variable}, stating whether
adversity bills a quota. The protocol applies on top of existing benchmarks,
not in place of them. Three directions remain open: a controlled ownership decomposition on a
benchmark we do not own, which requires showing that reassigning decisions
preserves task semantics; extending the audit beyond what a zero-call probe
battery can reach, so that scorers requiring model calls to replay are
assessed rather than recorded as \emph{not probed}; and protocol variants that
lift the L2 interface floor so that de-scaffolded scoring extends to larger
record sets.

\bibliography{references}

\clearpage
\appendix
\renewcommand{\topfraction}{0.9}
\renewcommand{\bottomfraction}{0.6}
\renewcommand{\textfraction}{0.05}
\renewcommand{\floatpagefraction}{0.75}
\setcounter{topnumber}{4}\setcounter{bottomnumber}{3}\setcounter{totalnumber}{6}
\renewcommand{\dbltopfraction}{0.9}
\renewcommand{\dblfloatpagefraction}{0.75}
\setcounter{dbltopnumber}{3}
\setcounter{figure}{0}\renewcommand{\thefigure}{A\arabic{figure}}
\setcounter{table}{0}\renewcommand{\thetable}{A\arabic{table}}
\section*{Supplementary Material}

This supplement provides the extended analyses referenced in the main text and
the complete experimental record. Unless noted otherwise, every table and
figure is regenerated directly from the released result files by
\texttt{gen\_appendix.py}.

\section{Extended Results and Analysis}
\label{sec:supp-extended}

This section holds the full version of every passage the main text
abbreviates. Nothing is dropped: all results, tables, and figures from the
project appear here or in the appendices.

\subsection{Pipeline and Scaffolding Spectrum (Figure)}
\label{sec:supp-framework}
\begin{figure*}[t]\centering
\includegraphics[width=0.92\textwidth]{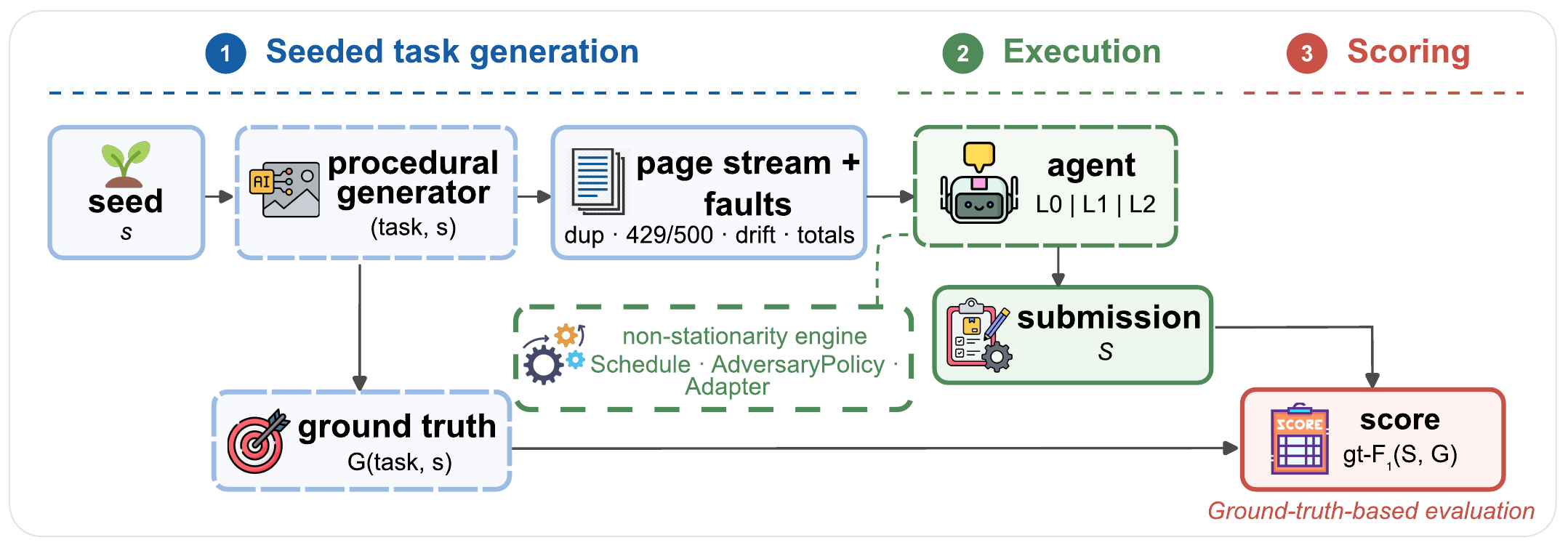}
\caption{The \bench{} pipeline in three stages.
(1)~\emph{Seeded task generation} (blue): a seed
drives the procedural generator, which emits the canonical ground-truth
record set $G$ and serves a fault-injected page stream (duplicates,
429/500, page drift, totals rows); (2)~\emph{execution} (green): the
agent, at scaffolding level L0--L2, produces a submission $S$, with the
non-stationarity engine (\texttt{Schedule}, \texttt{AdversaryPolicy},
\texttt{EnvironmentAdapter}) at the agent--environment boundary;
(3)~\emph{scoring} (red): $S$ is scored as \gtf{} against $G$.
The measured per-model scores at each scaffolding level are in
Figure~\ref{fig:scaffolding-spectrum}.}\label{fig:pipeline}
\end{figure*}

Figure~\ref{fig:pipeline} gives the one-page visual summary of the
measurement pipeline; the measured per-model scores at each scaffolding
level (defined in the main text, Section~4.2) are in
Figure~\ref{fig:scaffolding-spectrum}.

\begin{figure}[ht]\centering
\includegraphics[width=\columnwidth]{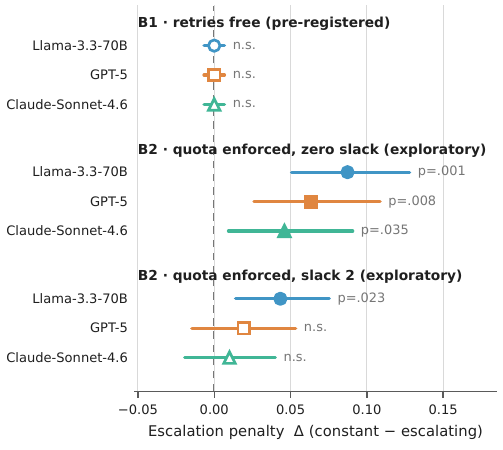}
\caption{The escalation penalty $\Delta$ (positive = escalating trajectory
hurts beyond matched-mean constant stress) with 95\% bootstrap CIs, $n{=}20$
paired seeds per arm. Under free retries (B1, pre-registered) the penalty is
zero for all three models. Adding one mechanism (an enforced request quota
that every fetch consumes, squeezed mid-episode only in the escalating arm)
makes the penalty positive on all three uncorrected (B2); after Holm
correction it survives for Llama and GPT-5 only, and a fixed no-LLM
always-retry policy achieves $\Delta = +0.096$, above every model. With
post-squeeze slack, the penalty persists only for Llama and vanishes for the
skip-hedging Claude and GPT-5. Filled markers: CI excludes 0; hollow:
n.s. All escalation experiments (B1, B2, the slack and dose ablations, and
the degradation curves) use exactly the three models named in the frozen
pre-registration; Claude-Fable-5 postdates the freeze and is absent from
the B-series by design, not omission.}\label{fig:escalation}
\end{figure}

\subsection{B1: Instrument Deviation and Power Study (full)}
\label{sec:supp-b1}
The confirmatory B1 run used the substrate \texttt{T9\_clean\_ab} under the
matched-mean A/B design, three models (Llama-3.3-70B, GPT-5,
Claude-Sonnet-4.6), retries free. It deviated from the frozen
pre-registration, and per the pre-registration's own rule (any deviation is
reported as exploratory) we disclose it: the frozen document specified the
de-scaffolded slim agent with the terminal judge reward as the outcome,
whereas the run used the L1 half-scaffold agent with ground-truth fetch
coverage. The substituted instrument isolates the retry/skip decision and is
therefore structurally blind to the pre-registered dedup/totals effect
channel: coverage deduplicates fetched records into a set and filters totals
rows before scoring, so duplicate injection and totals contamination cannot
move the outcome at all, and free retries neutralize the 429 faults. The only
channel through which a model could affect its score was an explicit SKIP.
The pre-registered decision rule fires only if the paired bootstrap CI
excludes 0 on the positive side \emph{and} a sign-flip permutation test gives
$p < 0.05$. The result is null for every model: $\Delta = +0.000$, 95\% CI
$[-0.007, +0.007]$, $p = 1.0$, paired Cliff's $\delta = +0.05$, 95\% CI
$[-0.30, +0.40]$. The interval is tight against large effects only, and
dominance beyond $\pm 0.4$ is excluded. SKIP, meanwhile, was chosen in 0 of
120 episodes, so the one model-sensitive channel never fired at all. That is
not a powered null, and we do not report it as one.

A Monte-Carlo power study (400 replications) shows the design detects a large
effect ($\Delta = 0.209$, the magnitude of the prior ``T9 gap'') with power
$0.998$ at $n = 20$, already $0.92$ at $n = 12$, while holding size at
$0.04$--$0.09$. Medium ($\Delta = 0.058$) and small effects are not excluded.
The simulation lets fragility act directly on the outcome, so the power figure
certifies the statistical test and says nothing about whether the instrument
could have picked up a behavioral effect. That prior gap (21 judge points, the
$\Delta = 0.209$ above) was a confound
(scaffolding plus a single seed plus a reasoning model running slow against a
step budget), not a non-stationarity effect. Under free retries the always-retry policy makes both arms' per-seed outcomes
identical across models within each arm, fixed by the seeded fault sequence rather than by
anything a model did.

We additionally ran the pre-registration-faithful instrument on all three
pre-registered models (slim agent, terminal judge reward, identical
adversary; $n{=}20$/arm), and report it under the frozen decision rule
without post-hoc adjustment. On the terminal judge reward Llama gives
$\Delta = +0.005$ ($p = .386$, $\delta = +0.05\,[-0.40,+0.45]$) and Claude
$\Delta = +0.016$ ($p = .167$, $\delta = +0.35\,[-0.05,+0.70]$). Both are
null. GPT-5 gives $\Delta = +0.011$ with a bootstrap CI $[+0.002, +0.023]$
excluding zero, sign-flip $p = .034$, and $\delta = +0.25\,[-0.15,+0.65]$.
The pre-registration's primary outcome was ``$\geq 1$ frontier reasoning
model,'' so under its own rule this is a \emph{confirmation}. The ground-truth
scorer accounts for it:
on the identical episodes GPT-5's and Claude's \gtf{} deltas are
$-4{\times}10^{-5}$ and its coverage delta is $\approx 0$
($1.1{\times}10^{-17}$), and Llama's deltas are null though noisier
(coverage $\Delta = -0.047$, $p = .75$). The submitted data is equally
correct across arms (identical unique-row, duplicate, and totals counts), so
the pre-registered confirmation lives \emph{entirely} in judge dimensions
that ground truth does not measure. The escalation effect the
pre-registration set out to find is real under the scorer it named and
vanishes under ground truth: the scorer-dependence this paper diagnoses,
surfacing inside our own confirmatory experiment
(\texttt{results\_b1\_slim.json}).

\begin{table}[t]
\centering
\small
\setlength{\tabcolsep}{1.6pt}
\resizebox{\columnwidth}{!}{%
\begin{tabular}{@{}lcccccc c@{}}
\toprule
Model & $\Delta$ & 95\% CI & Cliff's $\delta$ [CI] & $p$ & $p_{\mathrm{Holm}}$ & Deaths & Skips \\
\midrule
\multicolumn{8}{@{}l}{\emph{B1: free retries (pre-registered), $n{=}20$/arm}} \\
All three models  & $+0.000$ & $[-0.007,0.007]$ & $+0.05\,[-0.30,0.40]$ & $1.00$  & n/a     & n/a   & 0\,/\,0 \\
\midrule
\multicolumn{8}{@{}l}{\emph{B2: enforced quota 8, squeeze $-3$ (exploratory)}} \\
Llama-3.3-70B     & $+0.087$ & $[0.051,0.128]$  & $+0.55\,[0.25,0.80]$  & $.0011$ & $.0066$ & 11/20 & 6\,/\,1 \\
GPT-5             & $+0.063$ & $[0.026,0.109]$  & $+0.25\,[-0.10,0.60]$ & $.0076$ & $.0380$ & 8/20  & 9\,/\,4 \\
Claude-Sonnet-4.6 & $+0.046$ & $[0.009,0.091]$  & $+0.00\,[-0.40,0.40]$ & $.0355$ & $.1065$ & 7/20  & 11\,/\,7 \\
\midrule
\multicolumn{8}{@{}l}{\emph{B2-slack: quota 10, post-squeeze slack 2 (exploratory)}} \\
Llama-3.3-70B     & $+0.043$ & $[0.014,0.075]$  & $+0.25\,[-0.10,0.60]$ & $.0231$ & $.0924$ & 2/20  & 5\,/\,0 \\
GPT-5             & $+0.019$ & $[-0.015,0.053]$ & $+0.10\,[-0.25,0.50]$ & $.3495$ & $.6990$ & 1/20  & 6\,/\,3 \\
Claude-Sonnet-4.6 & $+0.010$ & $[-0.019,0.040]$ & $+0.00\,[-0.35,0.40]$ & $.6031$ & $.6990$ & 0/20  & 7\,/\,5 \\
\bottomrule
\end{tabular}}
\caption{Escalating vs.\ matched-mean constant stress. $\Delta$ $=$ paired
mean coverage difference (constant $-$ escalating; positive $=$ escalation
penalty), with seeded paired-bootstrap 95\% CI and sign-flip permutation $p$.
Cliff's $\delta$ $=$ within-pair dominance
$[\#(\Delta_i{>}0)-\#(\Delta_i{<}0)]/n$ with seeded bootstrap 95\% CI: only
Llama's B2 penalty is a dominance effect; Claude's nominally significant one
is magnitude-driven ($\delta = 0.00$, equal wins and losses).
$p_{\mathrm{Holm}}$ $=$ Holm-adjusted $p$ over the full six-test exploratory
family (B2 $+$ B2-slack; B1 is pre-registered and outside the family): only
Llama's and GPT-5's B2 penalties survive correction.
Deaths $=$ escalating-arm quota exhaustions; Skips $=$ SKIP actions,
escalating\,/\,constant arm. B1's rows are identical for all three models.}
\label{tab:escalation}
\end{table}

\subsection{B2 Rule-Agent Calibration (full table)}
\label{sec:supp-rulecal}
\begin{table}[ht]
\centering
\small
\setlength{\tabcolsep}{2.2pt}
\begin{tabular}{@{}lccccc@{}}
\toprule
Rule policy & $\Delta$ & 95\% CI & $p$ & Skips & Deaths \\
\midrule
\multicolumn{6}{@{}l}{\emph{Zero slack (quota 8, squeeze $-3$)}} \\
always-retry & $+0.096$ & $[0.061,0.137]$   & $.0004$ & 0  & 12 \\
quota-aware  & $+0.063$ & $[0.026,0.109]$   & $.0076$ & 12 & 9 \\
always-skip  & $-0.014$ & $[-0.055,0.027]$  & $.5726$ & 34 & 4 \\
\midrule
\multicolumn{6}{@{}l}{\emph{Slack 2 (quota 10, squeeze $-3$)}} \\
always-retry & $+0.015$ & $[-0.003,0.039]$  & $.3017$ & 0  & 2 \\
quota-aware  & $+0.027$ & $[-0.009,0.063]$  & $.2062$ & 9  & 2 \\
always-skip  & $-0.014$ & $[-0.055,0.027]$  & $.5726$ & 35 & 0 \\
\bottomrule
\end{tabular}
\caption{No-LLM rule-agent calibration on the identical B2 protocol
($n{=}20$ paired seeds; skips and quota deaths are totals over both arms'
40 episodes). At zero slack the fixed always-retry policy reproduces the
flip ($\Delta = +0.096$, exceeding every model in
Table~\ref{tab:escalation}), and the deterministic quota-aware rule is
bit-identical to GPT-5's summary; at slack 2 no rule policy is significant.}
\label{tab:rulecal}
\end{table}

\subsection{Zero-Slack Decision-Invariance and the Slack Regime (full)}
\label{sec:supp-slack}
\textbf{What the flip does not show.}
At zero post-squeeze slack (quota $8 - 3 = 5$ for 5 pages), retry and skip are
outcome-equivalent (each fault costs one page either way), so once the squeeze
lands, the escalating arm is decision-invariant by construction. The data
confirms it: the escalating-arm per-seed coverage vectors are \emph{identical}
for Llama and GPT-5, and Claude's differs on exactly 1 of 20 seeds (all
vectors are in the released JSON). The escalation penalty in this
configuration is therefore \emph{resource arithmetic, not decision failure}:
no high-coverage policy could have avoided it (the always-skip rule avoids it
in the $\Delta$ sense, $\Delta = -0.014$, n.s., but only by depressing
coverage in both arms, $0.704/0.690$ against the $0.83$ ceiling).

\textbf{Where slack exists, models differ in a counterintuitive direction.}
The constant arm never triggers the squeeze and keeps 3 pages of slack, so
retrying is affordable there; and that is where the models come apart. Llama
plays it near-optimally: 1 skip in 20 episodes, coverage $0.823$, near the
$0.83$ duplicate-overwrite ceiling. Claude skips 7 times \emph{unnecessarily}
and pays for it ($0.774$): the scarcity framing alone, with the budget never
actually at risk (0 constant-arm quota deaths), induces over-conservatism in
the frontier model. The slack ablation (quota 10, so 2 pages of post-squeeze
slack) flips the value of that same trait. Claude's skip-hedging now buys
insurance: zero quota deaths, and the escalation penalty vanishes
($\Delta = +0.010$, n.s.). GPT-5 likewise escapes it ($\Delta = +0.019$,
CI $[-0.015, +0.053]$, $p = .3495$, n.s., one quota death). Llama keeps a
nominally significant penalty ($\Delta = +0.043$, uncorrected $p = .0231$;
$p_{\mathrm{Holm}} = .0924$, 2 quota deaths). The rule calibration
re-attributes that persistence: at slack 2 the always-retry rule is \emph{not}
significant ($\Delta = +0.015$, $p = .30$; Table~\ref{tab:rulecal}), so
retrying alone no longer carries a penalty. Llama's constant arm matches the
always-retry rule exactly ($0.832$), while its escalating arm ($0.789$ against
the rule's $0.817$) is degraded by its five chosen SKIPs; the persistent
penalty is skip-induced over-conservatism under escalating scarcity, not
retry-heaviness. Skip propensity is thus a cost under abundance and a hedge under escalating
scarcity, and only its timing decides which. Neither policy dominates.

\begin{figure}[ht]\centering
\includegraphics[width=\columnwidth]{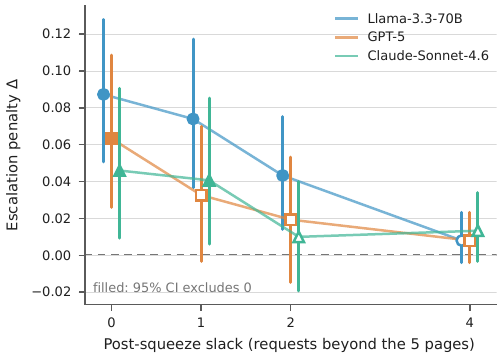}
\caption{Dose response: escalation penalty $\Delta$ (95\% bootstrap CI)
versus post-squeeze slack, all three models at $n{=}20$/arm per point
(the pre-registered trio; Claude-Fable-5 postdates the freeze and is
absent from the B-series by design).
Filled markers: CI excludes 0. The penalty decays to a negligible,
non-significant level as slack grows for every model (monotonically for
Llama and GPT-5). Reported as estimation; the confirmatory family is
unchanged.}
\label{fig:dose}
\end{figure}

\textbf{Dose response (full curve).} Two further budget configurations
(quota 9 and 12, i.e.\ post-squeeze slack 1 and 4; $n{=}20$/arm each) complete
a dose-response curve over slack (Figure~\ref{fig:dose}). We report these as
\emph{estimation}, not additional confirmatory tests. The penalty decays with
slack for all three models, monotonically for Llama
$+0.087 \to +0.074 \to +0.043 \to +0.008$ (slack $0 \to 1 \to 2 \to 4$) and
GPT-5 $+0.063 \to +0.033 \to +0.019 \to +0.008$, and to a negligible level
for Claude $+0.046 \to +0.041 \to +0.010 \to +0.013$ (whose slack-2 and
slack-4 points are statistically indistinguishable). Llama's CI excludes zero
through slack 2; GPT-5's and Claude's include zero from slack 2 onward. At
slack 4 the penalty is negligible for every model; with four spare requests
nothing binds, so decisions are once more outcome-free. Decision relevance is
thus an \emph{inverted U} in slack: at zero slack every choice is forced, at
high slack no choice matters, and models can be told apart only in the regime
between.

\subsection{Frontier Within-Model Fragility and the T8 Mixed-Fault Case}
\label{sec:supp-frontier}
The reliability metrics are also a within-model discipline for the frontier
tier. GPT-5 and Claude-Sonnet score $F_1 = 1.0$ on dedup and both retry
tasks but
drop to $0.80$ on page-drift, a recall failure (recall $0.67$): the models
emit clean, deduplicated rows but miss a third of the canonical set when page
contents are non-deterministic (Claude-Fable-5 clears page-drift at
$1.000$). The mixed-fault task T8, now complete for all eight models and
part of main-text Figure 2, separates the frontier further.
GPT-5's per-seed record is bimodal at the top,
$[1,1,1,1,0,.81,1,1,.81,.81]$: six perfect episodes, three at $0.81$, and
one full-episode zero (mean $0.844$, worst $0.000$), handing the 3rd-ranked
model by mean the suite's largest reliability gap ($0.930$).
Claude-Fable-5 and Claude-Sonnet-4.6 show the complementary profile,
never a perfect T8 episode (means $0.869$/$0.914$) but never a collapse
(worst $0.824$/$0.824$), while Claude-Haiku-4.5, GPT-4o, and GPT-4o-mini
sit at a flat $0.824$ (recall $0.70$, precision $1.00$, near-deterministic
across seeds). Consistent but capped and sometimes perfect but brittle are distinct
reliability profiles at a similar mean.

\subsection{T7: the L2 Re-Emission Ceiling, Measured}
\label{sec:supp-t7}
T7 (totals trap) scales the totals-row fault family to $750$ canonical
rows served at page size $250$. A solvability audit shows that under the
L2 protocol this cell is not a fair reliability measurement. The slim agent
must re-emit every cleaned row through a single \texttt{submit\_results}
call, and one row costs $\approx 54$ tokens of output. The budgets we grant
($6{,}000$ tokens non-reasoning, $12{,}000$ reasoning) therefore cap recall at
$\approx 0.15$/$0.30$ and $F_1$ at $\approx 0.26$/$0.46$ \emph{regardless of
model ability}. The
measured outcomes sit far below even those caps and collapse for every
model run: GPT-4o-mini $0.000$, GPT-5 $0.017$, Claude-Haiku $0.023$,
GPT-4o $0.026$, Claude-Fable-5 $0.043$, Claude-Sonnet $0.127$ (all
$n{=}10$; per-seed vectors in Table~\ref{tab:e1seeds}), with a
characteristic high-precision/near-zero-recall signature (precision
$\approx 1.0$). We therefore report T7 as a \emph{measured scale bound of de-scaffolded
scoring} rather than as a Table-1 cell. It marks the point where the L2
re-emission demand stops measuring execution decisions and starts measuring
output budgets. It is the mechanical-floor channel that floors Llama at small
$N$, now binding every model at large $N$. Protocol
variants that lift the bound (chunked submission, id-only re-emission) are
deferred to future work; Llama and Qwen were not run on T7 (their L2
floor at $N \le 60$ already binds).

\subsection{Model Access and Decoding Disclosure}
\label{sec:supp-endpoints}
All E1 episodes call provider APIs through one OpenAI-compatible client
against pinned endpoints: \texttt{api.anthropic.com/v1} (model ids
\texttt{claude-fable-5}, \texttt{claude-sonnet-4-6},
\texttt{claude-haiku-4-5}), \texttt{api.openai.com/v1} (\texttt{gpt-5},
\texttt{gpt-4o}, \texttt{gpt-4o-mini}), and \texttt{api.together.xyz/v1}
(\texttt{Llama-3.3-70B-Instruct-Turbo},
\texttt{Qwen2.5-7B-Instruct-Turbo}); rows were collected June--July 2026
(the T7/T8 completion runs on 2026-07-08 through 2026-07-10). The client
requests \texttt{temperature}$\,=0$ and an output budget of $6{,}000$
tokens ($12{,}000$ for the reasoning-class Claude-Fable-5, GPT-5, and
Claude-Sonnet-4.6). Two provider quirks, pinned by tests in the released
client, qualify that setting: GPT-5 rejects the \texttt{temperature} and
\texttt{stop} parameters (and takes \texttt{max\_completion\_tokens}), and
Claude-Fable-5's endpoint rejects \texttt{temperature} as deprecated for
that model, so those two models sample at their providers' defaults
rather than at a client-pinned $0$, and their per-seed variance
(Table~\ref{tab:e1seeds}) should be read with that in mind. All other
models run at \texttt{temperature}$\,=0$. Per-model scores are
point-in-time measurements of these dated endpoints, as the main-text
Limitations states.

\subsection{GPT-4o Bimodality: Full Mechanism}
\label{sec:supp-forensics-mech}
The worst-case reorder rests on GPT-4o's genuine bimodality on dedup (T3):
per-seed $F_1 = [0,0,1,0,0,0,0,1,1,1]$ at $n{=}10$, extended to $14/30$ solved
at $n{=}30$ (Wilson 95\% CI $[0.30, 0.64]$; Appendix~\ref{sec:forensics}).
Every episode opens with the same off-by-one call,
\texttt{fetch\_page(page=0)} against the 1-indexed API, deterministically
rejected with HTTP 422. All 14 solves recover by probing page 1 after three or four rejections. The
16 failures re-issue the byte-identical call four to nine times; the 422
carries the same \texttt{retry:\,true} flag as a genuine transient fault, so
the repeated call is consistent with the retry signal the model receives. Each then \emph{knowingly} submits a
schema-valid payload with zero data rows and a run log stating the data could
not be fetched (recall $=$ precision $=$ $0.00$). The failures thus end in the same
empty-payload channel that floors Llama and Qwen, entered upstream: rows never
fetched, rather than fetched and not re-emitted. The outcome is fully
determined by whether the model ever varies the page argument ($14/14$ of
solves, $0/16$ of failures): the bimodality is identifiable error-recovery
behavior in the transcript, not mysterious seed-to-seed variance. What is
model-dependent is the \emph{frequency} of entering that channel under an
identical demand (GPT-4o escapes on 14 of 30 runs, the floored tier on none),
so the worst-case reorder rests on error-recovery reliability, not on the
dedup decisions themselves: every episode that obtained rows deduplicated
them perfectly. GPT-4o-mini shows the identical terminal signature one tier
down (bimodal on page-drift).

\subsection{External Scorer Probe: Method and a Trivial-Agent Finding}
\label{sec:supp-external}
We probed the official scorer of $\tau$-bench \citep{yao2024tau} with zero
model calls, replaying its released historical trajectories through its
released reward function (\texttt{external/taubench\_probe.py}) and perturbing
only the semantics. A verbatim passing trajectory scores $1.0$; the same
trajectory with one database write semantically corrupted (shape preserved)
scores $0.0$; a trajectory whose required answer string is wrong (shape
preserved) scores $0.0$. $\tau$-bench's scorer is outcome-based (it hashes
the final database state) and therefore does \emph{not} carry the
shape-scorer artifact: the confound's two halves are separable design
properties, not a universal condition. The \emph{scaffold} half, however, is
present as an uncontrolled axis: the released harness ships four selectable
scaffolds whose differences are material and silent: the tool-calling
scaffold truncates parallel tool calls to the first and treats malformed
tool-call JSON as episode-fatal, while the ReAct scaffold forgives the same
error. The probe additionally surfaces a trivial-agent artifact even in this
outcome-based benchmark: $5$ of $115$ $\tau$-bench retail test tasks are
satisfied by an empty ``do-nothing'' trajectory, because their ground truth
requires no database write and no output, an outcome-based analogue of the
empty-submission floor we document on our own judge.

\subsection{External Scaffold-Variance Slice: Full Matrix}
\label{sec:supp-taubench-slice}
We measured scaffold variance on $\tau$-bench directly: three models from
two providers (Claude-Sonnet-4.6, Claude-Haiku-4.5, GPT-4o)
$\times$ all four scaffolds the harness ships, 15 retail test tasks per
cell (task ids 0--14, one trial), with the LLM user simulator held fixed
(Claude-Haiku-4.5) and \texttt{temperature}$\,=0$. No new code: only the
\texttt{-{}-agent-strategy} flag changes between cells.

\begin{table}[ht]\centering\small
\begin{tabular}{lccc}
\toprule
Scaffold & Sonnet-4.6 & Haiku-4.5 & GPT-4o \\
\midrule
tool-calling & 0.733 & 0.667 & 0.667 \\
few-shot     & 0.733 & 0.533 & 0.600 \\
act          & 0.667 & 0.600 & 0.667 \\
react        & 0.533 & 0.467 & 0.400 \\
\bottomrule
\end{tabular}
\caption{$\tau$-bench retail, average reward over the same 15 tasks per
cell. Changing only the scaffold flag moves the same model by up to
$0.267$ (GPT-4o); react is the worst scaffold for all three models; and
few-shot triples the Sonnet--Haiku gap. The slice measures
scaffold variance \emph{within} a model, not model ranking; the user
simulator is Claude-Haiku-4.5 throughout, itself one of the evaluated
agents (disclosed).}
\label{tab:taubench-slice}
\end{table}

\begin{figure}[ht]\centering
\includegraphics[width=\columnwidth]{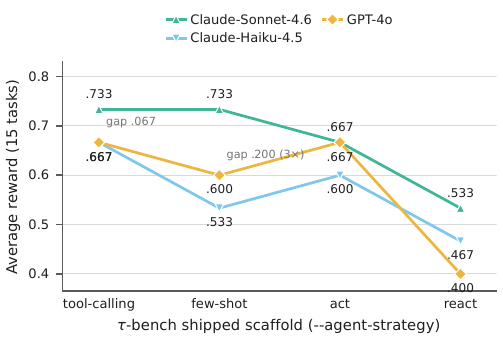}
\caption{The $\tau$-bench scaffold--model interaction:
Table~\ref{tab:taubench-slice} as lines. Each line is one agent across the
four shipped scaffolds (same 15 tasks, fixed user simulator); react is the
worst scaffold for all three models, the within-model spread reaches
$0.267$ (GPT-4o), and few-shot triples the Sonnet--Haiku gap
($0.067 \to 0.200$) while GPT-4o crosses between the two: the induced
ranking is scaffold-dependent. Recomputed from the
released per-episode JSONs by \texttt{Figures/figure8\_interaction.py}.}
\label{fig:tb-interaction}
\end{figure}

Four observations. (1)~\emph{Absolute scores do not transfer across
scaffolds}: Sonnet spans $0.733$--$0.533$, Haiku $0.667$--$0.467$, and
GPT-4o $0.667$--$0.400$, a swing of up to $0.267$ (4 of 15 tasks) for
the identical model, tasks, and user; react is the worst scaffold for all
three models.
(2)~\emph{Ordering is scaffold-dependent in general}: across tool-calling,
act, and react the Sonnet--Haiku contrast is monotone and rank-preserving
(Sonnet ahead by exactly one task at each level), but few-shot is Sonnet's
joint-best scaffold and Haiku's second-worst (a scaffold--model
interaction that triples that gap), and GPT-4o crosses between the two:
it ties Haiku under tool-calling, ties Sonnet under act, sits between them
under few-shot, and finishes last under react. The apparent ranking therefore depends on
which scaffold the leaderboard fixes. (3)~\emph{Statistical footing}:
per-model
paired sign tests between the extreme scaffolds do not individually reach
significance at $n{=}15$ (Sonnet 4~up/1~down, $p=.375$; Haiku 5~up/2~down,
$p=.453$; GPT-4o 6~up/2~down, $p=.289$); the slice is reported as
estimation, with its strength in the
cross-cell consistency rather than any single contrast.
(4)~\emph{Test--retest}: a full rerun of the (Sonnet, tool-calling) cell
under the identical configuration scores $0.867$ against the original
$0.733$. Thirteen of 15 tasks agree, and both flips, tasks 4 and 14, are
$0 \to 1$. Single-trial noise on this cell ($0.133$) is therefore of the same
order as the cross-scaffold spread, which is why we report the slice as
estimation rather than as a hypothesis test. Raw result files,
per-task pairings, the rerun
(\texttt{retest\_tool-calling-\allowbreak claude-sonnet-4-6\_0709155107.json}), and the
exact run commands are in
\texttt{external/taubench\_slice/}.
Figure~\ref{fig:tb-interaction} plots the matrix as an interaction chart,
showing the within-model spread, the react floor, and the few-shot split.

One caveat on the few-shot column. \citet{kapoor2025hal} report that
$\tau$-bench's official few-shot agent loads demonstration data that
overlaps the benchmark's own test set, and exclude that scaffold from
their analysis; the file they identify is in the airline domain, while
our slice is retail. The headline swing of $0.267$ (GPT-4o, tool-calling
versus react) does not pass through few-shot and is unaffected. We record
the issue because the few-shot cells, and the Sonnet--Haiku gap we read
off them, should be interpreted with it in mind.

\subsection{The Non-Stationarity Engine (full)}
\label{sec:supp-engine}

The main text summarizes the engine; here is the full architecture. To
generate stress reproducibly, and to test whether its \emph{shape}
matters, a domain-agnostic engine decouples three concerns.
\texttt{Schedule} maps an \texttt{AgentProgress} snapshot to an
intensity $i \in [0,1]$: the shape of the non-stationarity. Intensity is
driven by the agent's \emph{success} count, not elapsed steps (linear,
step, and plateau schedules).
\texttt{AdversaryPolicy} maps progress to a \texttt{Perturbation},
an \emph{intent}: a transient fault, an observation corruption
(duplicate/reorder/noise/drop), a budget delta, or a latency. The reference
\texttt{EscalatingAdversary} sets fault probability to
$\text{base} + \text{scale} \cdot i$ and corruption rate to
$\text{scale} \cdot i$; it knows nothing about trade data.
\texttt{EnvironmentAdapter}, the only domain-coupled piece,
realizes an intent in one environment: \texttt{ComtradeAdapter} produces 429
responses, injected duplicate rows, and a totals-row distractor once intensity
crosses \texttt{totals\_at}. The \texttt{AdversarialWrapper} ties them at the
agent--environment boundary, appending per-step intensity to a trace the
metrics consume; a fault cancels a success, so escalation tracks real
progress. Reproducibility is built into the framework: per-step randomness is
\texttt{Random(sha256("\{seed\}:\{step\}"))}, so the same seed yields a
bit-for-bit identical adversarial sequence.

\textbf{Controlled stress for Claim B.} The engine supplies the matched-mean
A/B design the main text pre-registers. The substrate is a
clean task (75 rows over 5 pages, own fault mode disabled), so the wrapper is
the sole source of adversity and the arms differ only in difficulty
\emph{trajectory}: the escalating arm ramps $0 \to 1.0$; the constant
arm holds the ramp's mean, $0.60$.
Both apply 429 faults and duplicate injection at matched mean, and both
inject totals-row contamination once intensity $\geq 0.70$, a threshold
the escalating arm crosses late and the constant arm never reaches: the
nonlinearity through which the pre-registration expected the two matched-mean
arms to diverge. The confirmatory run's coverage outcome
neutralizes this channel by construction, a deviation disclosed and
analyzed with the B1 results above.

\subsection{Agent-Layer Evaluations (full)}
\label{sec:supp-agentlayers}

The three optional LLM-assisted layers of the pipeline are measured, not
assumed. A scaffold-discovery step that proposes ownership maps from
harness source classifies 13 to 14 of 15 decisions correctly in each of
three independent runs against this paper's three declared reference
harnesses (the recurrent miss is the one genuinely ambiguous cell). A
synthesis step that generates a de-scaffolded harness variant passes
compilation, an independent discovery cross-audit (all five decisions
model-owned), and a live ground-truth-scored episode. A probe-design
step, given only a scorer's source, predicts the scorer kind for all
three audited scorers and independently reproduces the human probe batteries at
3 of 4 to 4 of 4 coverage under a fixed keyword rubric, and adds probes the
human batteries lacked. A narration layer is constrained by a number
guard: any figure in generated prose that does not appear in the audit
JSON is rejected.

\subsection{The Full Validity Cards}
\label{sec:supp-cards}

Table~\ref{tab:card-full} is the complete per-benchmark validity card;
the main text summarizes the licensed-claim lattice. Every cell is pass, fail, or
\emph{not probed}, with the evidence inline. The columns disagree in
different directions: the audit discriminates conditions rather than
finding the same defect everywhere.

\paragraph{Decision rules.} Each row is populated deterministically from the
audit artifacts under a fixed rule. \emph{Criterion sensitivity} passes iff the
scorer strictly separates ground-truth quality on the forged battery, i.e.\
$j(\text{correct}) > j(x) + \varepsilon$ for every submission $x$ of strictly
lower ground-truth quality (empty, fabricated, duplicate-laden), with
$\varepsilon$ the scorer's resolution; it fails on any tie or inversion across a
material quality gap (\bench{}: $j(\text{fabricated}) = j(\text{correct})$, gap
$0$) and is \emph{not probed} when no canonical-correctness oracle can be
constructed. \emph{Ownership completeness} passes iff
$D_{\mathrm{claimed}} \subseteq D_{\mathrm{model}}$ and every claimed decision is
non-degenerate ($\operatorname{Var}(A) > 0$ with outcome sensitivity); it fails
when a claimed execution-critical decision is harness-owned or degenerate
($\tau$-bench: reward moves $0.267$ across shipped scaffolds; \bench{} L1: SKIP
$0/120$) and is \emph{not probed} when the harness is not inspectable (BFCL:
single-turn decoding scaffold). \emph{Replay} passes iff runs reproduce
bit-for-bit from $(\text{task}, \text{seed})$. The \emph{licensed claim} is the
output row, and it follows a fixed lattice:

\begin{itemize}\setlength\itemsep{0pt}
\item \textbf{model capability} when replay, criterion sensitivity, and
      ownership completeness all pass;
\item \textbf{model--scaffold system performance} when criterion passes but
      ownership fails;
\item \textbf{self-report compliance} when the scorer is shape-based, so
      criterion fails;
\item \textbf{specification compliance} when a specification scorer passes and
      ownership is not probed.
\end{itemize}

\noindent A not-probed row never licenses the corresponding claim.

\begin{table*}[!tp]
\centering
\footnotesize
\setlength{\tabcolsep}{4pt}
\renewcommand{\arraystretch}{1.12}
\begin{tabular}{@{}l>{\raggedright\arraybackslash}p{4.7cm}>{\raggedright\arraybackslash}p{4.4cm}>{\raggedright\arraybackslash}p{3.8cm}@{}}
\toprule
Audit item & \bench{} (as shipped) & $\tau$-bench (retail slice) & BFCL (AST slice) \\
\midrule
Model-sensitive ownership & \textbf{fail}: L0 owns 5/5 decisions; submissions byte-identical, 21/21 model pairs (7 models, 3 providers) & \textbf{fail}: reward moves 0.267 across the 4 shipped scaffolds; ranking scaffold-dependent & not probed \\
Criterion-aligned scorer & \textbf{fail}: fabricated $=$ correct ($0.980$), gt $F_1$ 0~vs~1 & \textbf{pass}: corrupt DB write $\to 0.0$ (aligned to final DB state) & \textbf{pass}: spec-based, 0/2000 deviate \\
Null floor & \textbf{fail}: empty submission scores $0.645$ & \textbf{pass}: no shape floor observed & \textbf{pass}: empty arguments fail \\
Trace preference & \textbf{fail}: misreported correct data $-0.398$ & not probed & not probed (no trace channel) \\
Decision non-degeneracy & \textbf{fail}: L1 SKIP 0/120 (always-retry) & not probed & not probed \\
Interface floor & \textbf{fail}: L2 re-emission floor (floored tier) & not probed & not probed \\
Joint intervention & separation only at L2$\,\times\,$gt (main-text Table~1) & not probed (scorer swap unnecessary) & not probed \\
Reliability reporting & \textbf{pass}: 8 models, 50--70 seeded episodes each & est.\ only: 15 distinct tasks/cell, one trial each (not seeded repeats) & not probed (scorer-only audit) \\
\midrule
Valid interpretation & L0 $=$ system; L2$+$gt $=$ model execution within measured interface bounds & per model--scaffold configuration & spec-valid calls \\
Invalid interpretation & shipped leaderboard as model capability & single-scaffold ranking as model ranking & spec pass $=$ execution success \\
Overall verdict & repair required; model-valid after repair, within measured interface bounds & system-valid, scaffold-sensitive & scorer spec-valid; scaffold unaudited \\
\bottomrule
\end{tabular}
\caption{Full validity cards for the three audited benchmarks (numbers
from the released BenchAudit artifacts). \emph{Not probed} is a
first-class cell value: an unaudited item is never reported as passing.
The judge-probe cells come from the audit run on the 25-row dedup task;
Table~\ref{tab:judgeprobe} puts the same three canned submissions through the
same scorer on \texttt{T2\_multi\_page} (2{,}345 rows) and reports $0.987$ and
$0.648$. The two runs differ only in the observability dimension, and the
fabricated-equals-correct tie holds at both sizes.}
\label{tab:card-full}
\end{table*}

\paragraph{Audit coverage and cost.} All three targets are audited by one
key-free pipeline that makes \emph{zero} model calls: it regenerates every
\bench{} card cell from the seeded core, replays $\tau$-bench's released reward
function and scaffold grid, and runs BFCL's vendored AST checker offline.
Table~\ref{tab:coverage} reports what each target exercises. Integrating a target
costs benchmark-specific code (\bench{}: \texttt{ComtradeAdapter}, 111 lines;
$\tau$-bench: 196; BFCL: 174) against a shared, benchmark-agnostic core (probe
battery, joint intervention, reliability, and card synthesis; ${\approx}1000$
lines). The complete run finishes in seconds, and its 13 built-in oracle checks
(6 for \bench{}, 4 for $\tau$-bench, 3 for BFCL) all pass, so the auditor is
itself an executable regression suite. $\tau$-bench and BFCL are integrated as
benchmark-specific audit modules feeding the shared card synthesizer, not full
re-implementations of the ownership intervention: $\tau$-bench's ownership
evidence is scaffold variance (not a controlled de-scaffolding) and BFCL's
scaffold is not probed. A controlled ownership decomposition on $\tau$-bench is
future work.

\begin{table*}[t]
\centering
\small
\setlength{\tabcolsep}{6pt}
\begin{tabular}{@{}l>{\raggedright\arraybackslash}p{7.4cm}cc@{}}
\toprule
Target & Stages audited & Oracle checks & Model calls \\
\midrule
\bench{} & scorer probe, $2{\times}2$ joint intervention, seeded reliability, ownership map, card & 6/6 & 0 \\
$\tau$-bench (retail) & scorer-probe replay, scaffold variance & 4/4 & 0 \\
BFCL (AST slice) & scorer probe (AST checker) & 3/3 & 0 \\
\bottomrule
\end{tabular}
\caption{BenchAudit cross-target coverage from the key-free audit run (zero
model calls; each stage completes in under a second). \emph{Oracle checks} are
the pipeline's built-in regression assertions; all pass, reproducing every
reported number.}
\label{tab:coverage}
\end{table*}

\paragraph{Target interface.} A benchmark is audited through four key-free
hooks, which \texttt{ComtradeAdapter} implements for \bench{}:

\begin{itemize}\setlength\itemsep{0pt}
\item a \emph{scaffold-ownership map}, assigning each execution-critical
      decision (retry, dedup, totals-drop, pagination, submission) to harness
      or model at each level. It is declared in v0, and agentically discovered
      then human-confirmed in v1;
\item a \emph{task registry} enumerating the seeded tasks;
\item a \emph{canonical-truth oracle} returning ground truth from
      $(\text{task},\text{seed})$;
\item a \emph{scorer hook} that replays the benchmark's own scorer on a
      submission.
\end{itemize}

\noindent Table~\ref{tab:iface} shows how the two external targets supply
these hooks.

\begin{table*}[t]
\centering
\small
\setlength{\tabcolsep}{5pt}
\begin{tabular}{@{}l>{\raggedright\arraybackslash}p{3.3cm}lll@{}}
\toprule
Hook & Consumed by & \bench{} & $\tau$-bench & BFCL \\
\midrule
Ownership map & ownership audit, $2{\times}2$ & declared L0/L1/L2 & 4 shipped scaffolds (uncontrolled) & not probed \\
Task registry & all stages & seeded generator & retail slice & AST slice \\
Canonical-truth oracle & ground-truth scorer, criterion probe & seeded generator & final DB state & annotated spec \\
Scorer hook & criterion probe, $2{\times}2$ & shipped judge & released reward fn & AST checker \\
\bottomrule
\end{tabular}
\caption{The four target hooks and how each audited benchmark supplies them.
A missing hook yields \emph{not probed} for the stages that consume it, never a
pass.}
\label{tab:iface}
\end{table*}

\paragraph{Failure modes.} When a hook is unavailable or a model-owned channel
is degenerate, the audit does not fabricate a verdict; it records \emph{not
probed} or \emph{fail} by the Decision-rules above. Table~\ref{tab:failmodes}
enumerates the resolutions.

\begin{table*}[t]
\centering
\small
\setlength{\tabcolsep}{6pt}
\begin{tabular}{@{}l>{\raggedright\arraybackslash}p{7.2cm}@{}}
\toprule
Condition & Audit response \\
\midrule
No canonical-truth oracle & criterion probe $=$ not probed (metamorphic tests only) \\
No scorer access & scorer rows $=$ not probed \\
Harness not inspectable & ownership $=$ not probed (BFCL) \\
Ownership map declared but unverified & flagged for human confirmation (stage 1) \\
Non-deterministic replay & replay/reliability $=$ fail \\
Model-owned decision degenerate ($\operatorname{Var}=0$) & ownership $=$ fail (\bench{} L1: SKIP 0/120) \\
De-scaffolding alters task semantics & intervention not licensed (equivalence unestablished) \\
\bottomrule
\end{tabular}
\caption{Failure modes and the audit's deterministic response. \emph{Not probed}
and \emph{fail} are first-class outcomes; an unmet condition is never reported as
a pass.}
\label{tab:failmodes}
\end{table*}

\section{Extended External Audit: a Scorer-Kind Census Across Nine
Benchmarks}
\label{sec:supp-census}

The main text audits three benchmarks (Section~5.4). This section widens that
audit to nine, to ask how far the two halves of the confound generalize.
Everything here is offline and makes zero model calls: scorer classifications
are grounded in each benchmark's released scorer source, and executed
deviation counts are reported only where the scorer was actually run on a
canned battery; source-level classifications are labeled as such.

We classify each shipped scorer into one of four kinds. \emph{Shape /
self-report}: grades output format, field presence, or self-reported metadata,
never the submitted values against ground truth (a well-formed fabrication can
score as high as the truth). \emph{Outcome}: grades the final answer or world
state against ground truth (test pass/fail, final DB state, answer match,
produced-file content). \emph{Specification}: grades conformance to a required
spec (e.g.\ AST match), certifying validity rather than execution success.
\emph{LLM-judge}: an LLM decides pass/fail or win/lose; it grades
\emph{plausibility} rather than ground truth (in the audited case no gold
answer exists at all), so a confident, well-formed fabrication can be labeled
as passing, and every verdict requires a model call, so this kind is
\emph{not} offline-probeable.

\begin{table*}[t]\centering\small
\setlength{\tabcolsep}{6pt}
\begin{tabular}{@{}llccl@{}}
\toprule
Benchmark & Primary scorer kind & Fabricated $=$ correct? & Offline-probeable? & Evidence \\
\midrule
\bench{} & shape / self-report & \textbf{yes} & yes & executed \\
$\tau$-bench \citep{yao2024tau} & outcome (DB state) & no & yes & executed \\
BFCL \citep{patil2025bfcl} & specification (AST) & no (0/2000) & yes & executed \\
GAIA \citep{mialon2024gaia} & outcome (answer) & no (0/5) & yes & executed \\
SWE-bench \citep{jimenez2024swebench} & outcome (tests) & no & partial & source \\
WebArena \citep{zhou2024webarena} & outcome $+$ llm-judge & no (partial) & partial & source \\
AgentBench \citep{liu2023agentbench} & mostly outcome & mostly no & partial & source \\
OSWorld \citep{xie2024osworld} & outcome (state) & partial & partial & source \\
ToolBench \citep{qin2023toolllm} & \textbf{llm-judge} & \textbf{yes} & \textbf{no} & source \\
\bottomrule
\end{tabular}
\caption{Scorer-kind census. \emph{Fabricated $=$ correct}: can a fabricated
but well-formed submission score as high as a correct one? \emph{Offline-probeable}:
can the scorer run as a pure function on canned submissions with zero model
calls (``partial'': the score math is pure but acquiring the end state needs a
live environment). \emph{Evidence}: ``executed'' $=$ a canned battery was run
through the released scorer; ``source'' $=$ classification from the released
scorer code, decisive lines inspected, deviation counts not run. Distribution
of primary kinds: shape 1, outcome 6, specification 1, llm-judge 1.}
\label{tab:census}
\end{table*}

Three observations follow. First, the \emph{shape} defect of Section~3 (main
text) is benchmark-specific, and the census quantifies this: exactly one of
nine benchmarks (\bench{}, as shipped) carries a pure shape / self-report
scorer, and most shipped scorers are value-grounded. As an executed
spot-check of one of them, we ran GAIA's released \texttt{question\_scorer}
(vendored, unmodified) on a canned battery over five gold-answer types:
correct submissions pass 5/5, formatting variants (whitespace, case,
\texttt{\$}/\texttt{\%} units) pass 5/5, and fabricated-but-well-formed values
pass 0/5.

Second, the census surfaces a fourth scorer kind beyond the three named in the
main text: \emph{llm-judge} (ToolBench's ToolEval as the primary case;
WebArena's fuzzy-match branch and one AgentBench environment as minor cases).
Shape scoring and llm-judging are two routes to the same criterion-validity
failure: neither anchors the score to ground truth, so a well-formed
fabrication can pass. The main text diagnoses the first. The census adds the
second, which zero-call probes cannot reach at all.

Third, the census explains why the main text's $2{\times}2$ joint intervention
(Table~1, main text) is difficult to replicate externally. That design needs a
shape scorer \emph{and} a constructible ground truth in the same benchmark, and
the two rarely co-occur: value-grounded benchmarks lack the broken column,
while llm-judge benchmarks, which run against live services with no gold
answer, lack the truth column. \bench{}'s seeded generator supplies both, so it can serve as a controlled
testbed for the repair and not only as a diagnosis.

\section{Complete Experimental Record}
\label{sec:appendix}

This appendix carries the complete experimental record behind the aggregates
in the main text. Every data-bearing table below is generated directly from
the released result files or released source code by
\texttt{gen\_appendix.py}; nothing is transcribed by hand. The one exception
is the qualitative positioning matrix (Table~\ref{tab:posmatrix}), which
contains no measured numbers.

\subsection{Positioning Matrix}
The qualitative matrix promised in the related-work section. Cells are our
reading of each suite's released evaluation protocol, and ``partial'' entries
are marked deliberately in the prior suites' favor. $\tau$-bench's
pass$^k$ does report cross-trial reliability, though without seeded
worst-case or CVaR estimation over environment adversity. SWE-bench admits
multiple community scaffolds, so multi-scaffold comparisons exist, but the
scaffolding level is never a controlled experimental variable. Harness-Bench
varies complete harness configurations factorially across model backends and
explicitly disclaims mechanism-level decomposition, with one trajectory per
task--model--harness cell.

\begin{table*}[t]\centering\small
\setlength{\tabcolsep}{4pt}
\begin{tabular}{lp{3.7cm}p{4.2cm}p{4.2cm}}
\toprule
Suite & Adversity location & Worst-case / cross-seed reliability &
Scaffolding level as controlled variable \\
\midrule
ToolBench \citep{qin2023toolllm} & prompt/task specification & no (mean pass
rate) & no \\
BFCL \citep{yan2024berkeley} & none (static calls) & no (mean accuracy) &
no \\
API-Bank \citep{li2023apibank} & none (static dialogues) & no (mean
accuracy) & no \\
AgentBench \citep{liu2023agentbench} & task environments, not fault
injection & no (mean success) & no \\
$\tau$-bench \citep{yao2024tau} & rule/policy level & \emph{partial}:
pass$^k$ over repeated trials; no seeded worst-case or CVaR & no \\
SWE-bench \citep{jimenez2024swebench} & none (static issues) & no (mean
resolve rate) & \emph{partial}: community scaffolds vary
\citep{yang2024sweagent,xia2024agentless}, but the level is not an
experimental axis \\
Harness-Bench \citep{yao2026harnessbench} & none (sandboxed offline
workflow tasks) & no (one trajectory per task--model--harness cell) &
\emph{partial}: six complete harnesses varied factorially across eight
backends with native behavior preserved; configuration-level diagnosis, not
an ownership axis \\
\midrule
\bench{} & environment response dynamics (seeded fault injection) &
worst-case, CVaR@$0.2$, Reliability@$\tau$ over seeded adversary draws &
yes: ordinal L0/L1/L2 ownership axis \\
\bottomrule
\end{tabular}
\caption{Positioning matrix over the seven suites cited in the related-work
section. Qualitative; no measured numbers.}
\label{tab:posmatrix}
\end{table*}

\subsection{Judge Probe: Three Canned Submissions Through the
Released Scorer}
The direct verification of the shape-scoring claim
(Section 3 (main text)). \texttt{probe\_judge.py} passes three canned
submissions for \texttt{T2\_multi\_page} through the \emph{released}
\texttt{judge.py}, unmodified: (a)~an empty \texttt{data.jsonl} with
well-formed, self-consistent metadata and a disciplined run log; (b)~2{,}345
fabricated rows with the correct schema, plausible values, self-consistent
metadata, and invented record ids (zero overlap with the seeded ground
truth); (c)~the exact seeded ground-truth record set, emitted by the
benchmark's own generator. All three cases carry the \emph{identical} run
log and metadata shape; only the submitted record set (and the truthful
declared \texttt{row\_count}) varies. The fabricated and correct submissions
receive the same judge total and the same six-dimension breakdown; only
\gtf{} separates them.

\begin{table*}[t]\centering\small
\begin{tabular}{lrrrrrrrrr}
\toprule
Submission & Rows & Corr./30 & Compl./15 & Rob./15 & Eff./15 & DQ/15 & Obs./10 & Judge total & \gtf{} \\
\midrule

empty & 0 & 18.00 & 15.00 & 15.00 & 7.50 & 5.00 & 4.33 & 0.648 & 0.000 \\
fabricated & 2345 & 30.00 & 15.00 & 15.00 & 15.00 & 15.00 & 8.67 & 0.987 & 0.000 \\
correct & 2345 & 30.00 & 15.00 & 15.00 & 15.00 & 15.00 & 8.67 & 0.987 & 1.000 \\

\bottomrule
\end{tabular}
\caption{Judge probe on \texttt{T2\_multi\_page}: six-dimension breakdown (points), judge total ($/100$, normalized to $[0,1]$), and \gtf{} of the same submission, from \texttt{probe\_judge\_results.json}. The empty case trips the correctness governance gate ($<70\%$), which halves its efficiency and observability credit; it still floors at $0.648$. The fabricated and correct record sets are indistinguishable to the judge ($0.987$ each) and maximally separated by \gtf{} ($0.0$ vs.\ $1.0$).}
\label{tab:judgeprobe}
\end{table*}

\subsection{Per-Task Parameters}
The complete parameter set of the ten benchmark tasks plus the clean A/B base
task, read directly from the released \texttt{server/tasks.py}. Every page of
data is a deterministic function of \texttt{(task\_id, seed)}; the fault
column lists the injection mode with its mode-specific parameters.

\begin{table*}[t]\centering\small
\begin{tabular}{llrrrrp{6.4cm}}
\toprule
Task & Paging & Page size & Max req. & Rows & QPS & Fault injection \\
\midrule

T1\_single\_page & page & 1000 & 5 & 800 & 5 & none \\
T2\_multi\_page & page & 500 & 50 & 2345 & 5 & pagination \\
T3\_duplicates & offset & 10 & 50 & 25 & 5 & duplicates: duplicate\_rate=0.08, cross\_page\_duplicate\_rate=0.03 \\
T4\_rate\_limit\_429 & page & 10 & 60 & 30 & 3 & rate\_limit: fail\_on=[2] \\
T5\_server\_error\_500 & page & 10 & 60 & 30 & 3 & server\_error: fail\_on=[2] \\
T6\_page\_drift & page & 12 & 60 & 36 & 5 & page\_drift \\
T7\_totals\_trap & offset & 250 & 60 & 750 & 5 & totals\_trap \\
T8\_mixed\_faults & page & 20 & 80 & 60 & 2 & mixed: rate\_limit\_fail\_on=[2, 5], duplicate\_rate=0.1, cross\_page\_duplicate\_rate=0.05 \\
T9\_adaptive\_adversary & page & 15 & 100 & 75 & 2 & adaptive: initial\_duplicate\_rate=0.05, escalation\_per\_page=0.03, adaptive\_429\_after\_page=3, adaptive\_totals\_after\_page=4 \\
T10\_constrained\_budget & page & 10 & 50 & 80 & 5 & duplicates: duplicate\_rate=0.05, cross\_page\_duplicate\_rate=0.02 \\
T9\_clean\_ab & page & 15 & 100 & 75 & 2 & none \\

\bottomrule
\end{tabular}
\caption{Per-task parameters, generated from the released \texttt{server/tasks.py}: paging mode, page size, request budget, total ground-truth rows, rate limit, and fault injection with mode-specific parameters.}
\label{tab:taskparams}
\end{table*}

\subsection{E1: Complete Per-Task Statistics, Including Diagnostic Runs}
The main-text Figure 2 scopes aggregates to the five
fully run execution tasks (T3--T6, T8), $n{=}10$ seeds per cell for all
eight models. The complete record below additionally includes the
T7 re-emission-ceiling diagnostic (Section~\ref{sec:supp-t7}; excluded
from aggregates because its 750-row re-emission demand exceeds
single-submission output budgets, capping $F_1$ at $0.26$/$0.46$ for
non-reasoning/reasoning token limits) and the partial T10 runs, with the
per-cell seed count $n$ disclosed. Metric: \gtf{} per episode; CVaR at
$\alpha{=}0.2$; Reliability at $\tau{=}0.5$.

\begin{table*}[t]\centering\small
\begin{tabular}{llrrrrrr}
\toprule
Model & Task & $n$ & Mean & Std & Worst & CVaR & Rel@$0.5$ \\
\midrule

Claude-Fable-5 & T3 dedup & 10 & 1.000 & 0.000 & 1.000 & 1.000 & 1.000 \\
Claude-Fable-5 & T4 429-retry & 10 & 1.000 & 0.000 & 1.000 & 1.000 & 1.000 \\
Claude-Fable-5 & T5 500-retry & 10 & 1.000 & 0.000 & 1.000 & 1.000 & 1.000 \\
Claude-Fable-5 & T6 page-drift & 10 & 1.000 & 0.000 & 1.000 & 1.000 & 1.000 \\
Claude-Fable-5 & T8 mixed & 10 & 0.869 & 0.030 & 0.824 & 0.824 & 1.000 \\
Claude-Fable-5 & T7 totals & 10 & 0.043 & 0.041 & 0.011 & 0.012 & 0.000 \\
GPT-5 & T3 dedup & 10 & 1.000 & 0.000 & 1.000 & 1.000 & 1.000 \\
GPT-5 & T4 429-retry & 10 & 1.000 & 0.000 & 1.000 & 1.000 & 1.000 \\
GPT-5 & T5 500-retry & 10 & 1.000 & 0.000 & 1.000 & 1.000 & 1.000 \\
GPT-5 & T6 page-drift & 10 & 0.806 & 0.017 & 0.800 & 0.800 & 1.000 \\
GPT-5 & T8 mixed & 10 & 0.844 & 0.294 & 0.000 & 0.406 & 0.900 \\
GPT-5 & T10 budget & 10 & 1.000 & 0.000 & 1.000 & 1.000 & 1.000 \\
GPT-5 & T7 totals & 10 & 0.017 & 0.032 & 0.000 & 0.000 & 0.000 \\
Claude-Sonnet-4.6 & T3 dedup & 10 & 1.000 & 0.000 & 1.000 & 1.000 & 1.000 \\
Claude-Sonnet-4.6 & T4 429-retry & 10 & 1.000 & 0.000 & 1.000 & 1.000 & 1.000 \\
Claude-Sonnet-4.6 & T5 500-retry & 10 & 1.000 & 0.000 & 1.000 & 1.000 & 1.000 \\
Claude-Sonnet-4.6 & T6 page-drift & 10 & 0.800 & 0.000 & 0.800 & 0.800 & 1.000 \\
Claude-Sonnet-4.6 & T8 mixed & 10 & 0.914 & 0.074 & 0.824 & 0.824 & 1.000 \\
Claude-Sonnet-4.6 & T7 totals & 10 & 0.127 & 0.059 & 0.005 & 0.059 & 0.000 \\
Claude-Haiku-4.5 & T3 dedup & 10 & 1.000 & 0.000 & 1.000 & 1.000 & 1.000 \\
Claude-Haiku-4.5 & T4 429-retry & 10 & 1.000 & 0.000 & 1.000 & 1.000 & 1.000 \\
Claude-Haiku-4.5 & T5 500-retry & 10 & 1.000 & 0.000 & 1.000 & 1.000 & 1.000 \\
Claude-Haiku-4.5 & T6 page-drift & 10 & 0.799 & 0.004 & 0.787 & 0.793 & 1.000 \\
Claude-Haiku-4.5 & T8 mixed & 10 & 0.824 & 0.000 & 0.824 & 0.824 & 1.000 \\
Claude-Haiku-4.5 & T7 totals & 10 & 0.023 & 0.045 & 0.000 & 0.000 & 0.000 \\
GPT-4o & T3 dedup & 10 & 0.400 & 0.490 & 0.000 & 0.000 & 0.400 \\
GPT-4o & T4 429-retry & 10 & 1.000 & 0.000 & 1.000 & 1.000 & 1.000 \\
GPT-4o & T5 500-retry & 10 & 1.000 & 0.000 & 1.000 & 1.000 & 1.000 \\
GPT-4o & T6 page-drift & 10 & 0.800 & 0.000 & 0.800 & 0.800 & 1.000 \\
GPT-4o & T8 mixed & 10 & 0.824 & 0.000 & 0.824 & 0.824 & 1.000 \\
GPT-4o & T10 budget & 2 & 0.000 & 0.000 & 0.000 & 0.000 & 0.000 \\
GPT-4o & T7 totals & 10 & 0.026 & 0.017 & 0.000 & 0.000 & 0.000 \\
GPT-4o-mini & T3 dedup & 10 & 0.000 & 0.000 & 0.000 & 0.000 & 0.000 \\
GPT-4o-mini & T4 429-retry & 10 & 0.900 & 0.300 & 0.000 & 0.500 & 0.900 \\
GPT-4o-mini & T5 500-retry & 10 & 1.000 & 0.000 & 1.000 & 1.000 & 1.000 \\
GPT-4o-mini & T6 page-drift & 10 & 0.327 & 0.268 & 0.000 & 0.000 & 0.600 \\
GPT-4o-mini & T8 mixed & 10 & 0.824 & 0.000 & 0.824 & 0.824 & 1.000 \\
GPT-4o-mini & T7 totals & 10 & 0.000 & 0.000 & 0.000 & 0.000 & 0.000 \\
Llama-3.3-70B & T3 dedup & 10 & 0.000 & 0.000 & 0.000 & 0.000 & 0.000 \\
Llama-3.3-70B & T4 429-retry & 10 & 0.000 & 0.000 & 0.000 & 0.000 & 0.000 \\
Llama-3.3-70B & T5 500-retry & 10 & 0.000 & 0.000 & 0.000 & 0.000 & 0.000 \\
Llama-3.3-70B & T6 page-drift & 10 & 0.000 & 0.000 & 0.000 & 0.000 & 0.000 \\
Llama-3.3-70B & T8 mixed & 10 & 0.000 & 0.000 & 0.000 & 0.000 & 0.000 \\
Llama-3.3-70B & T10 budget & 10 & 0.000 & 0.000 & 0.000 & 0.000 & 0.000 \\
Qwen2.5-7B & T3 dedup & 10 & 0.000 & 0.000 & 0.000 & 0.000 & 0.000 \\
Qwen2.5-7B & T4 429-retry & 10 & 0.000 & 0.000 & 0.000 & 0.000 & 0.000 \\
Qwen2.5-7B & T5 500-retry & 10 & 0.000 & 0.000 & 0.000 & 0.000 & 0.000 \\
Qwen2.5-7B & T6 page-drift & 10 & 0.000 & 0.000 & 0.000 & 0.000 & 0.000 \\
Qwen2.5-7B & T8 mixed & 10 & 0.000 & 0.000 & 0.000 & 0.000 & 0.000 \\

\bottomrule
\end{tabular}
\caption{E1 complete per-(model, task) statistics from \texttt{results\_reliability\_full.json}, including partial runs and uneven seed counts.}
\label{tab:e1full}
\end{table*}

\subsection{E1: Raw Per-Seed $F_1$ Vectors}
The seed-level grid behind Table~\ref{tab:e1full}. Solve-or-fail bimodality
(e.g.\ GPT-4o on T3) is directly visible.

\begin{table*}[t]\centering\footnotesize
\begin{tabular}{llp{11.5cm}}
\toprule
Model & Task & Per-seed \gtf{} \\
\midrule

Claude-Fable-5 & T3 dedup & [1, 1, 1, 1, 1, 1, 1, 1, 1, 1] \\
Claude-Fable-5 & T4 429-retry & [1, 1, 1, 1, 1, 1, 1, 1, 1, 1] \\
Claude-Fable-5 & T5 500-retry & [1, 1, 1, 1, 1, 1, 1, 1, 1, 1] \\
Claude-Fable-5 & T6 page-drift & [1, 1, 1, 1, 1, 1, 1, 1, 1, 1] \\
Claude-Fable-5 & T8 mixed & [0.823529, 0.888889, 0.823529, 0.888889, 0.888889, 0.888889, 0.823529, 0.888889, 0.888889, 0.888889] \\
Claude-Fable-5 & T7 totals & [0.0288714, 0.106061, 0.010596, 0.015852, 0.106061, 0.0132275, 0.106061, 0.015852, 0.015852, 0.015852] \\
GPT-5 & T3 dedup & [1, 1, 1, 1, 1, 1, 1, 1, 1, 1] \\
GPT-5 & T4 429-retry & [1, 1, 1, 1, 1, 1, 1, 1, 1, 1] \\
GPT-5 & T5 500-retry & [1, 1, 1, 1, 1, 1, 1, 1, 1, 1] \\
GPT-5 & T6 page-drift & [0.8, 0.8, 0.8, 0.8, 0.8, 0.8, 0.8, 0.8, 0.8, 0.857143] \\
GPT-5 & T8 mixed & [1, 1, 1, 1, 0, 0.811881, 1, 1, 0.811881, 0.811881] \\
GPT-5 & T10 budget & [1, 1, 1, 1, 1, 1, 1, 1, 1, 1] \\
GPT-5 & T7 totals & [0, 0, 0.0262812, 0, 0, 0, 0.103535, 0, 0, 0.0366492] \\
Claude-Sonnet-4.6 & T3 dedup & [1, 1, 1, 1, 1, 1, 1, 1, 1, 1] \\
Claude-Sonnet-4.6 & T4 429-retry & [1, 1, 1, 1, 1, 1, 1, 1, 1, 1] \\
Claude-Sonnet-4.6 & T5 500-retry & [1, 1, 1, 1, 1, 1, 1, 1, 1, 1] \\
Claude-Sonnet-4.6 & T6 page-drift & [0.8, 0.8, 0.8, 0.8, 0.8, 0.8, 0.8, 0.8, 0.8, 0.8] \\
Claude-Sonnet-4.6 & T8 mixed & [0.823529, 0.974359, 0.974359, 0.974359, 0.974359, 0.823529, 0.823529, 0.974359, 0.823529, 0.974359] \\
Claude-Sonnet-4.6 & T7 totals & [0.113208, 0.00531915, 0.192771, 0.113208, 0.148148, 0.113208, 0.113208, 0.113208, 0.113208, 0.24356] \\
Claude-Haiku-4.5 & T3 dedup & [1, 1, 1, 1, 1, 1, 1, 1, 1, 1] \\
Claude-Haiku-4.5 & T4 429-retry & [1, 1, 1, 1, 1, 1, 1, 1, 1, 1] \\
Claude-Haiku-4.5 & T5 500-retry & [1, 1, 1, 1, 1, 1, 1, 1, 1, 1] \\
Claude-Haiku-4.5 & T6 page-drift & [0.8, 0.8, 0.8, 0.8, 0.8, 0.8, 0.786885, 0.8, 0.8, 0.8] \\
Claude-Haiku-4.5 & T8 mixed & [0.823529, 0.823529, 0.823529, 0.823529, 0.823529, 0.823529, 0.823529, 0.823529, 0.823529, 0.823529] \\
Claude-Haiku-4.5 & T7 totals & [0, 0.113208, 0, 0, 0, 0.113208, 0, 0, 0, 0] \\
GPT-4o & T3 dedup & [0, 0, 1, 0, 0, 0, 0, 1, 1, 1] \\
GPT-4o & T4 429-retry & [1, 1, 1, 1, 1, 1, 1, 1, 1, 1] \\
GPT-4o & T5 500-retry & [1, 1, 1, 1, 1, 1, 1, 1, 1, 1] \\
GPT-4o & T6 page-drift & [0.8, 0.8, 0.8, 0.8, 0.8, 0.8, 0.8, 0.8, 0.8, 0.8] \\
GPT-4o & T8 mixed & [0.823529, 0.823529, 0.823529, 0.823529, 0.823529, 0.823529, 0.823529, 0.823529, 0.823529, 0.823529] \\
GPT-4o & T10 budget & [0, 0] \\
GPT-4o & T7 totals & [0, 0.0366013, 0, 0.0366013, 0, 0.0366013, 0.0366013, 0.0366013, 0.0366013, 0.0366013] \\
GPT-4o-mini & T3 dedup & [0, 0, 0, 0, 0, 0, 0, 0, 0, 0] \\
GPT-4o-mini & T4 429-retry & [0, 1, 1, 1, 1, 1, 1, 1, 1, 1] \\
GPT-4o-mini & T5 500-retry & [1, 1, 1, 1, 1, 1, 1, 1, 1, 1] \\
GPT-4o-mini & T6 page-drift & [0, 0, 0.530612, 0.530612, 0.530612, 0.615385, 0, 0.530612, 0, 0.530612] \\
GPT-4o-mini & T8 mixed & [0.823529, 0.823529, 0.823529, 0.823529, 0.823529, 0.823529, 0.823529, 0.823529, 0.823529, 0.823529] \\
GPT-4o-mini & T7 totals & [0, 0, 0, 0, 0, 0, 0, 0, 0, 0] \\
Llama-3.3-70B & T3 dedup & [0, 0, 0, 0, 0, 0, 0, 0, 0, 0] \\
Llama-3.3-70B & T4 429-retry & [0, 0, 0, 0, 0, 0, 0, 0, 0, 0] \\
Llama-3.3-70B & T5 500-retry & [0, 0, 0, 0, 0, 0, 0, 0, 0, 0] \\
Llama-3.3-70B & T6 page-drift & [0, 0, 0, 0, 0, 0, 0, 0, 0, 0] \\
Llama-3.3-70B & T8 mixed & [0, 0, 0, 0, 0, 0, 0, 0, 0, 0] \\
Llama-3.3-70B & T10 budget & [0, 0, 0, 0, 0, 0, 0, 0, 0, 0] \\
Qwen2.5-7B & T3 dedup & [0, 0, 0, 0, 0, 0, 0, 0, 0, 0] \\
Qwen2.5-7B & T4 429-retry & [0, 0, 0, 0, 0, 0, 0, 0, 0, 0] \\
Qwen2.5-7B & T5 500-retry & [0, 0, 0, 0, 0, 0, 0, 0, 0, 0] \\
Qwen2.5-7B & T6 page-drift & [0, 0, 0, 0, 0, 0, 0, 0, 0, 0] \\
Qwen2.5-7B & T8 mixed & [0, 0, 0, 0, 0, 0, 0, 0, 0, 0] \\

\bottomrule
\end{tabular}
\caption{E1 raw per-seed vectors (seed order 0,1,\dots).}
\label{tab:e1seeds}
\end{table*}

\subsection{Failure Forensics: GPT-4o on T3 Dedup at $n{=}30$}
\label{sec:forensics}
The T3 bimodality of the main text was re-run at $n{=}30$ with
the full message transcript, submitted payload, and judge breakdown captured
per episode (\texttt{harness\_gpt4o\_forensics.py} $\to$
\texttt{results\_gpt4o\_t3\_n30.json} and
\texttt{forensics\_gpt4o\_t3/seed\_NN.json}). The outcome is strictly
two-valued: 14/30 episodes at $F_1 = 1.0$ and
16/30 at $F_1 = 0.0$ (solve rate
$0.467$, Wilson 95\% CI $[0.302, 0.639]$,
consistent with the E1 estimate of $4/10$). The system and task prompts are
byte-identical across all 30 episodes, and every trajectory opens
with the identical call \texttt{fetch\_page(page=0)}, which the mock
API's 1-indexed paging (\texttt{page} $\geq 1$ validation)
deterministically rejects with HTTP 422, delivered in the same
\texttt{retry:\,true} envelope as the injected transient faults.
Table~\ref{tab:forensics} classifies every episode by its mechanically
checkable transcript signature; the classifier is re-run (and asserted) at
every regeneration of this appendix. No failing episode ever varies the
page argument, emits a malformed submission call, or claims success: each
re-issues the identical rejected call, announces it is giving up, and
submits a schema-valid payload with an empty data section, a truthful
\texttt{row\_count} of $0$, and a run log stating the data could not be
fetched. Conversely, no solved episode mishandles the task content: all
14 submit exactly the 25 unique ground-truth records
(duplicates removed, zero spurious rows).

\begin{table}[t]\centering\small
\setlength{\tabcolsep}{3pt}
\begin{tabular}{@{}p{6.0cm}rr@{}}
\toprule
Transcript signature & $n$ & Ex.\ seed \\
\midrule
Solved: probes page 1 after 3--4 rejections of page 0, fetches pages 1--3, submits 25 unique rows & 14 & 1 \\
Failed: 4 identical \texttt{page=0} calls, then schema-valid empty submission declaring fetch failure & 12 & 0 \\
Failed: extended perseveration (5--9 identical calls), same terminal capitulation & 4 & 17 \\
Failed: malformed submission, or success claimed without data & 0 & n/a \\
\bottomrule
\end{tabular}
\caption{Mechanism taxonomy over all 30 forensic episodes, computed by re-parsing every transcript in \texttt{forensics\_gpt4o\_t3/}. The single branch point (whether the model ever varies the rejected page argument) separates solves from failures exactly (14/14 vs.\ 0/16).}
\label{tab:forensics}
\end{table}

\subsection{B2 and B2-Slack: Raw Per-Seed Coverage and Decisions}
Escalating-arm coverage at zero slack is \emph{identical} for Llama and GPT-5
on all 20 seeds (Claude differs on seed 18 only): the decision-equivalence
disclosed in the main text, visible below rather than asserted. Decision
columns count model choices across the 20 episodes of each arm.

\begin{table*}[t]\centering\footnotesize
\begin{tabular}{llp{7.8cm}rrr}
\toprule
Model & Arm & Per-seed coverage & Retries & Skips & Quota deaths \\
\midrule

GPT-5 & escalating & [0.53, 0.68, 0.84, 0.81, 0.68, 0.68, 0.83, 0.68, 0.85, 0.83, 0.68, 0.71, 0.68, 0.71, 0.83, 0.68, 0.68, 0.83, 0.68, 0.84] & 8 & 9 & 8 \\
GPT-5 & constant@mean & [0.85, 0.83, 0.84, 0.83, 0.83, 0.65, 0.84, 0.67, 0.85, 0.81, 0.84, 0.83, 0.64, 0.85, 0.81, 0.85, 0.81, 0.83, 0.68, 0.84] & 13 & 4 & 0 \\
Claude-Sonnet-4.6 & escalating & [0.53, 0.68, 0.84, 0.81, 0.68, 0.68, 0.83, 0.68, 0.85, 0.83, 0.68, 0.71, 0.68, 0.71, 0.83, 0.68, 0.68, 0.83, 0.52, 0.84] & 6 & 11 & 7 \\
Claude-Sonnet-4.6 & constant@mean & [0.85, 0.83, 0.84, 0.83, 0.83, 0.65, 0.84, 0.67, 0.85, 0.81, 0.84, 0.83, 0.64, 0.68, 0.81, 0.85, 0.65, 0.83, 0.51, 0.84] & 10 & 7 & 0 \\
Llama-3.3-70B & escalating & [0.53, 0.68, 0.84, 0.81, 0.68, 0.68, 0.83, 0.68, 0.85, 0.83, 0.68, 0.71, 0.68, 0.71, 0.83, 0.68, 0.68, 0.83, 0.68, 0.84] & 11 & 6 & 11 \\
Llama-3.3-70B & constant@mean & [0.85, 0.83, 0.84, 0.83, 0.83, 0.81, 0.84, 0.83, 0.85, 0.81, 0.84, 0.83, 0.80, 0.85, 0.81, 0.85, 0.81, 0.83, 0.68, 0.84] & 16 & 1 & 0 \\

\bottomrule
\end{tabular}
\caption{B2 raw record, zero slack (budget $8\to5$), from \texttt{results\_b2\_squeeze.json}.}
\label{tab:b2raw}
\end{table*}

\begin{table*}[t]\centering\footnotesize
\begin{tabular}{llp{7.8cm}rrr}
\toprule
Model & Arm & Per-seed coverage & Retries & Skips & Quota deaths \\
\midrule

Claude-Sonnet-4.6 & escalating & [0.71, 0.68, 0.84, 0.81, 0.68, 0.83, 0.83, 0.68, 0.85, 0.83, 0.83, 0.85, 0.68, 0.85, 0.83, 0.83, 0.68, 0.83, 0.68, 0.84] & 12 & 7 & 0 \\
Claude-Sonnet-4.6 & constant@mean & [0.85, 0.83, 0.84, 0.83, 0.83, 0.65, 0.84, 0.67, 0.85, 0.81, 0.84, 0.83, 0.64, 0.85, 0.81, 0.85, 0.65, 0.83, 0.68, 0.84] & 12 & 5 & 0 \\
Llama-3.3-70B & escalating & [0.71, 0.68, 0.84, 0.81, 0.68, 0.83, 0.83, 0.68, 0.85, 0.83, 0.83, 0.85, 0.83, 0.85, 0.83, 0.83, 0.68, 0.83, 0.68, 0.84] & 16 & 5 & 2 \\
Llama-3.3-70B & constant@mean & [0.85, 0.83, 0.84, 0.83, 0.83, 0.81, 0.84, 0.83, 0.85, 0.81, 0.84, 0.83, 0.80, 0.85, 0.81, 0.85, 0.81, 0.83, 0.85, 0.84] & 17 & 0 & 0 \\

\bottomrule
\end{tabular}
\caption{B2 raw record, slack 2 (budget $10\to7$), from \texttt{results\_b2\_slack.json}.}
\label{tab:b2slack}
\end{table*}

\begin{table*}[t]\centering\footnotesize
\begin{tabular}{llp{7.8cm}rrr}
\toprule
Model & Arm & Per-seed coverage & Retries & Skips & Quota deaths \\
\midrule

GPT-5 & escalating & [0.71, 0.83, 0.84, 0.81, 0.68, 0.83, 0.83, 0.68, 0.85, 0.83, 0.83, 0.85, 0.68, 0.85, 0.83, 0.83, 0.68, 0.83, 0.68, 0.84] & 14 & 6 & 1 \\
GPT-5 & constant@mean & [0.85, 0.83, 0.84, 0.83, 0.83, 0.65, 0.84, 0.67, 0.85, 0.81, 0.84, 0.83, 0.64, 0.85, 0.81, 0.85, 0.81, 0.83, 0.85, 0.84] & 14 & 3 & 0 \\

\bottomrule
\end{tabular}
\caption{B2 raw record, slack 2 (budget $10\to7$), GPT-5 run, from \texttt{results\_b2\_slack\_gpt5.json}.}
\label{tab:b2slackgpt5}
\end{table*}

\subsection{Design Sensitivity: Monte-Carlo Power Study}
Simulated power of the paired matched-mean test by effect regime and seeds per
arm (400 replications per cell; paired bootstrap CI at $\alpha{=}0.05$
excluding zero as the decision rule). The pre-registered $n{=}20$ reaches
near-ceiling power for large effects and the null-control regime confirms the
test size.

\begin{table*}[t]\centering\small
\begin{tabular}{lrrrrrrrr}
\toprule
Regime & True $\Delta$ & $n{=}3$ & $n{=}5$ & $n{=}8$ & $n{=}12$ & $n{=}20$ & $n{=}30$ & $n{=}50$ \\
\midrule

null control & -0.000 & 0.08 & 0.07 & 0.07 & 0.05 & 0.09 & 0.06 & 0.04 \\
small effect & +0.032 & 0.11 & 0.12 & 0.10 & 0.14 & 0.18 & 0.26 & 0.31 \\
medium effect & +0.058 & 0.15 & 0.23 & 0.21 & 0.33 & 0.39 & 0.57 & 0.70 \\
large effect & +0.209 & 0.48 & 0.74 & 0.84 & 0.92 & 1.00 & 1.00 & 1.00 \\

\bottomrule
\end{tabular}
\caption{Power by effect regime and $n$ per arm.}
\label{tab:power}
\end{table*}

\subsection{Degradation Curves at Constant Intensity}
\begin{figure}[ht]\centering
\includegraphics[width=\columnwidth]{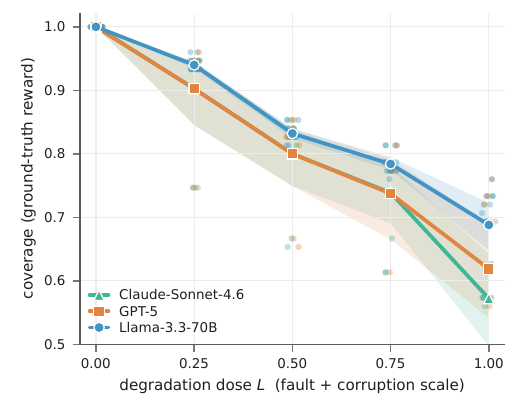}
\caption{Degradation dose-response: per-seed coverage (dots), the mean
(solid, with markers) and a seeded 95\% bootstrap CI band, versus constant
adversary dose under the quota-enforced half-scaffold instrument (no
mid-episode squeeze); three models, 10 seeds per point. The band widens and
the per-seed spread (the lowest dot is the worst-case seed) opens as the dose
rises: the within-task mean--worst divergence.}
\label{fig:degradation}
\end{figure}

Figure~\ref{fig:degradation} plots coverage against \emph{constant} adversary
dose under the quota-enforced half-scaffold instrument (budget 8, no
squeeze), 10 seeds per (model, level), from
\texttt{results\_degradation.json}. The per-seed spread below the mean opens
as the dose rises (a mean--worst gap of zero for every model at dose $0$,
$0.12$--$0.21$ at dose $1.0$): the within-task counterpart of the
suite-level mean--worst divergence in the main text.

\subsection{Pre-Registration Digest and Artifact Provenance}
The Claim~B protocol was frozen before the confirmatory run
(\texttt{PREREGISTRATION.md}): directional hypothesis (escalating worse),
design (matched-mean arms, $n{=}20$ paired seeds, three models), decision rule
(bootstrap CI excluding zero \emph{and} sign-flip $p<0.05$), and the explicit
deferral of the enforced-quota variant as an \emph{exploratory} ablation,
the variant reported here as B2.

\textbf{Disclosed deviation from the frozen document.} The pre-registration
specified the de-scaffolded slim agent with the terminal judge reward as the
outcome; the confirmatory run used the L1 half-scaffold agent with
ground-truth fetch coverage. Because coverage deduplicates fetched records
and filters totals rows before scoring, the pre-registered dedup/totals
effect channel has no causal path to the outcome, and with retries free the
only model-sensitive channel is an explicit SKIP, which fired in 0 of 120
episodes. Per the pre-registration's own rule, the run is therefore reported
as a null under an instrument that isolates a single decision, not as a
powered test of the frozen hypothesis (Section 5 (main text)); the
\texttt{"agent"} metadata field of \texttt{results\_t9\_ab\_full.json} has
been corrected from \texttt{slim} to \texttt{half\_scaffold\_coverage}. A
pre-registration-faithful rerun (slim agent, judge-reward outcome, identical
adversary; \texttt{harness\_b1\_slim.py} $\to$
\texttt{results\_b1\_slim.json}) has been completed for all three
pre-registered models and is reported in Section~5 (main text):
under the frozen decision rule GPT-5 confirms on the terminal judge reward
while its ground-truth and coverage deltas are $\approx 0$.
Artifact map:

\begin{itemize}\setlength\itemsep{0pt}
\item E1: \texttt{nonstationary/}\allowbreak\texttt{harness\_reliability\_full.py} $\to$
      \texttt{results\_reliability\_full.json}
\item B1 (pre-registered protocol, deviating instrument as disclosed above):
      \texttt{harness\_t9\_ab\_full.py} $\to$
      \texttt{results\_t9\_ab\_full.json}
\item B1 pre-registration-faithful rerun:
      \texttt{harness\_b1\_slim.py} $\to$ \texttt{results\_b1\_slim.json}
\item External scorer probe: \texttt{external/}\allowbreak\texttt{taubench\_probe.py} $\to$ \texttt{taubench\_probe\_results.json}
\item External scaffold-variance slice: \texttt{external/}\allowbreak\texttt{taubench\_slice/} (\texttt{run\_grid.sh}, \texttt{run\_fewshot.sh} $\to$ per-cell result JSONs; Table~\ref{tab:taubench-slice})
\item B2 / B2-slack (exploratory): \texttt{harness\_b2.py} $\to$
      \texttt{results\_b2\_squeeze.json}, \texttt{results\_b2\_slack.json},
      \texttt{results\_b2\_slack\_gpt5.json}
\item B2 rule-agent calibration (no-LLM bound):
      \texttt{harness\_b2\_rule.py} $\to$ \texttt{results\_b2\_rule.json}
\item GPT-4o T3 failure forensics ($n{=}30$, full transcripts;
      Section~\ref{sec:forensics}): \texttt{harness\_gpt4o\_forensics.py}
      $\to$ \texttt{results\_gpt4o\_t3\_n30.json},
      \texttt{forensics\_gpt4o\_t3/seed\_NN.json}
\item Degradation curves: \texttt{harness\_degradation.py} $\to$
      \texttt{results\_degradation.json}
\item T7/T8 completion runs: \texttt{harness\_reliability\_full.py} with
      \texttt{AB\_TASKS}/\texttt{AB\_MODELS} knobs plus the per-seed resumer
      \texttt{resume\_t7\_seeds.py} $\to$
      \texttt{results\_t7t8\_anthropic.json},
      \texttt{results\_t7t8\_openai.json}, merged by
      \texttt{merge\_t7t8.py} (refuses incomplete coverage) into
      \texttt{results\_reliability\_full.json}
\item Judge probe: \texttt{paper/aaai27/probe\_judge.py} $\to$
      \texttt{probe\_judge\_results.json} (Table~\ref{tab:judgeprobe}),
      against the released \texttt{server/judge.py} in the code release
\item Figures: \texttt{Figures/figure1\_regen.py} regenerates the
      pipeline/spectrum figure from
      \texttt{llm\_results\_\{llama,gpt5,claude\}.json},
      \texttt{results\_t9\_ab\_full.json}, and
      \texttt{results\_reliability\_full.json};
      \texttt{Figures/figure2.py} regenerates the
      escalation-forest figure from the B1/B2 files above;
      \texttt{Figures/figure4\_degradation.py} regenerates
      Figure~\ref{fig:degradation} from \texttt{results\_degradation.json};
      \texttt{Figures/figure6\_spectrum.py} regenerates the eight-model E1
      spectrum (Figure~\ref{fig:spectrum-eight}) from
      \texttt{results\_reliability\_full.json};
      \texttt{Figures/figure7\_judgeprobe.py} regenerates the judge-probe
      bars (Figure~\ref{fig:judgeprobe-bars}) from
      \texttt{probe\_judge\_results.json};
      \texttt{Figures/figure8\_interaction.py} regenerates the
      scaffold--model interaction chart from the per-episode JSONs in
      \texttt{external/taubench\_slice/};
      \texttt{Figures/figure9\_confound.py} draws the two-path confound
      schematic (no data inputs).
\end{itemize}
\section{Complete Figure Record}
\label{sec:appendix-figures}

This appendix is the complete figure record of the project. It has two parts.
First, the measurement figures: the scaffolding-spectrum figure, which places
the three models present at \emph{every} scaffolding level side by side
(plus Claude-Fable-5 at the one level it was run) and discloses the
provenance of every plotted number, and the eight-model E1 reliability
spectrum, the visual companion of main-text
Figure 2. Second, the legacy artifacts of the original
benchmark release: the judge-scored, production-scaffolded
(L0) charts and the GRPO~\citep{shao2024deepseekmath} training figures produced against the original
reward. Nothing here is re-scored or re-drawn; the legacy figures are
preserved exactly as released, because under the paper's diagnosis they are
primary evidence: their flat leaderboards are the double measurement
confound in action.

\subsection{The 2$\times$2 Joint Intervention}
\label{sec:fig-joint}
\begin{figure*}[!ht]
\centering
\includegraphics[width=\textwidth]{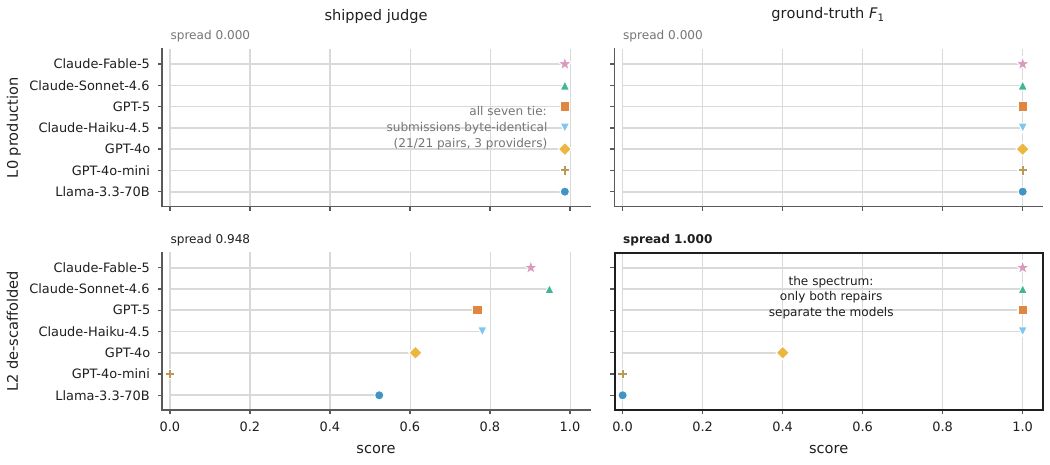}
\caption{The joint intervention of main-text Table 1 as a
picture: rows are the harness swap, columns are the scorer swap, and each
panel plots the same seven models on the same axis (values machine-read from
the released \texttt{joint\_matrix.json}). Top row: the shipped
condition; seven models from three providers collapse to one vertical
line under either scorer because the scaffold's submissions are
byte-identical. Bottom left: the judge separates only failure-to-submit
and misorders what it grades (Sonnet above the ground-truth-perfect
GPT-5). Bottom right: model-owned execution against ground truth
recovers the spectrum.}
\label{fig:joint}
\end{figure*}

Figure~\ref{fig:joint} draws main-text Table~1 as four panels, one for each
cell of the harness-by-scorer swap.

\subsection{The Scaffolding Spectrum}
\label{sec:fig-scaffolding}

\begin{figure}[ht]
\centering
\includegraphics[width=\columnwidth]{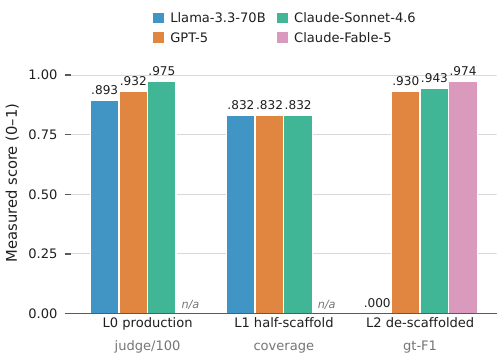}
\caption{The scaffolding spectrum: measured performance of the three models
present at every scaffolding level, plus Claude-Fable-5 at the one level it
was run (L2; its L0/L1 slots read \emph{n/a}, absence rather than zero).
The metric changes with the level,
and that is the point: L0 (production scaffold) is the original six-dimension
judge score, $/100$ normalized to $[0,1]$; L1 (half-scaffold) is coverage;
L2 (de-scaffolded) is \gtf{} averaged over the 50 runs of the five execution tasks. L0 and L1 are
flat or
near-flat across models; only L2 separates them. Bars compare models
\emph{within} a level; the three metrics are not commensurable \emph{across}
levels. Every value is read programmatically from the released result JSONs
by \texttt{Figures/figure3\_scaffolding.py}; nothing is hand-typed.}
\label{fig:scaffolding-spectrum}
\end{figure}

\begin{table}[t]\centering\small
\begin{tabular}{lrrr}
\toprule
Model & L0 (judge) & L1 (cov.) & L2 ($F_1$) \\
\midrule
Llama-3.3-70B     & 0.8932 & 0.8320 & 0.0000 \\
GPT-5             & 0.9320 & 0.8320 & 0.9299 \\
Claude-Sonnet-4.6 & 0.9750 & 0.8320 & 0.9428 \\
Claude-Fable-5    & n/a    & n/a    & 0.9739 \\
\bottomrule
\end{tabular}
\caption{Exact values plotted in
Figure~\ref{fig:scaffolding-spectrum} (model $\times$ scaffolding level):
L0 judge score $/100$, L1 coverage, L2 \gtf{} over the five execution tasks
(mean of the 50 per-run rewards). Claude-Fable-5 was run at L2 only.}
\label{tab:scaffolding-values}
\end{table}

Figure~\ref{fig:scaffolding-spectrum} plots the values in
Table~\ref{tab:scaffolding-values}. Provenance of each column:

\begin{itemize}\setlength\itemsep{0pt}
\item \textbf{L0 (production scaffold).} Mean of the 10 per-task
      \texttt{score} values, $/100$, from
      \texttt{llm\_results\_\{llama,gpt5,claude\}.json}. Each mean matches
      the file's own \texttt{average\_score} field (89.3, 93.2, 97.5).
\item \textbf{L1 (half-scaffold).} \texttt{summary.cm\_mean} (the constant
      arm) from \texttt{results\_t9\_ab\_full.json}. The escalating-arm mean
      is also 0.832 for all three models, so the mean over both arms is
      identical to the value shown.
\item \textbf{L2 (de-scaffolded).} Mean over the 50 per-run
      \texttt{rewards} of
      \texttt{T3\_duplicates}, \texttt{T4\_rate\_limit\_429},
      \texttt{T5\_server\_error\_500}, \texttt{T6\_page\_drift}, and
      \texttt{T8\_mixed\_faults} from
      \texttt{results\_reliability\_full.json}, the same five fully-run
      execution tasks scoped by main-text Figure 2
      (identical to its Mean column). Claude-Fable-5 has no L0/L1 entries:
      it was added to the suite at E1 (L2) time and never run under the
      production or half scaffold.
\end{itemize}

\noindent\textbf{Disclosed deviation from the figure brief.} The brief
anticipated L0 spanning roughly $0.93$--$0.975$, but Llama-3.3-70B's measured
L0 is $0.893$ (its released \texttt{average\_score} is 89.3). The measured
value is plotted unchanged; no number was adjusted to match the brief. The
figure was visually verified after generation (no clipping; axis, legend, and
per-level metric sublabels legible), and both \texttt{.pdf} and \texttt{.png}
renders are produced by the same script.

\subsection{The Judge Probe, Visualized}
\label{sec:fig-judgeprobe}
\begin{figure}[ht]
\centering
\includegraphics[width=\columnwidth]{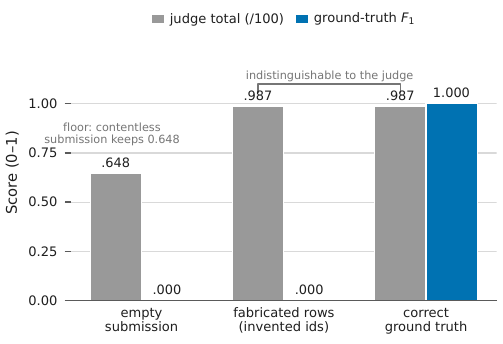}
\caption{The judge probe (Table~\ref{tab:judgeprobe}) as bars: three
canned submissions through the released, unmodified \texttt{judge.py},
each scored by the judge total ($/100$, gray) and by \gtf{} (blue). The
two scorer artifacts are visible at a glance: a contentless submission is
floored at $0.648$, and the fabricated record set is indistinguishable
from the correct one to the judge ($0.987$ each) while \gtf{} separates
them at $0.000$ versus $1.000$. Values read from
\texttt{probe\_judge\_results.json} by
\texttt{Figures/figure7\_judgeprobe.py}; nothing is hand-typed.}
\label{fig:judgeprobe-bars}
\end{figure}

Figure~\ref{fig:judgeprobe-bars} plots Table~\ref{tab:judgeprobe} as bars, so
the floor and the tie can be read off at a glance.

\subsection{The Eight-Model E1 Reliability Spectrum}
\label{sec:fig-spectrum}
\begin{figure}[ht]
\centering
\includegraphics[width=\columnwidth]{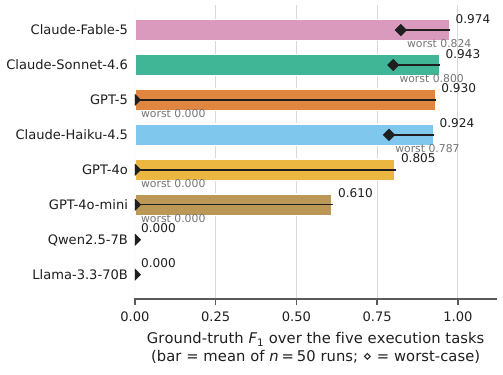}
\caption{The eight-model E1 reliability spectrum, the visual companion of
main-text Figure 2: mean \gtf{} (bar) and worst-case
over all 50 runs (diamond) per model at L2 on the five execution tasks,
sorted by mean. The
bar-to-diamond distance is the reliability gap. Claude-Fable-5 tops both
mean and worst case without saturating; GPT-5's gap ($0.93$, a mixed-fault
hard zero) is now the suite's largest: the worst-case reorder of the
main text, reaching the frontier. Every
value is computed from \texttt{results\_reliability\_full.json} by
\texttt{Figures/figure6\_spectrum.py} (mean over the 50 per-run rewards;
worst $=$ minimum); nothing is hand-typed, and the values match
main-text Figure 2 exactly.}
\label{fig:spectrum-eight}
\end{figure}

Figure~\ref{fig:spectrum-eight} is the visual companion of main-text Figure~2:
one bar and one worst-case marker per model.

\subsection{Legacy Artifacts: the Original Benchmark Release}
\label{sec:fig-legacy}

\begin{figure*}[t]
\centering
\includegraphics[width=0.9\textwidth]{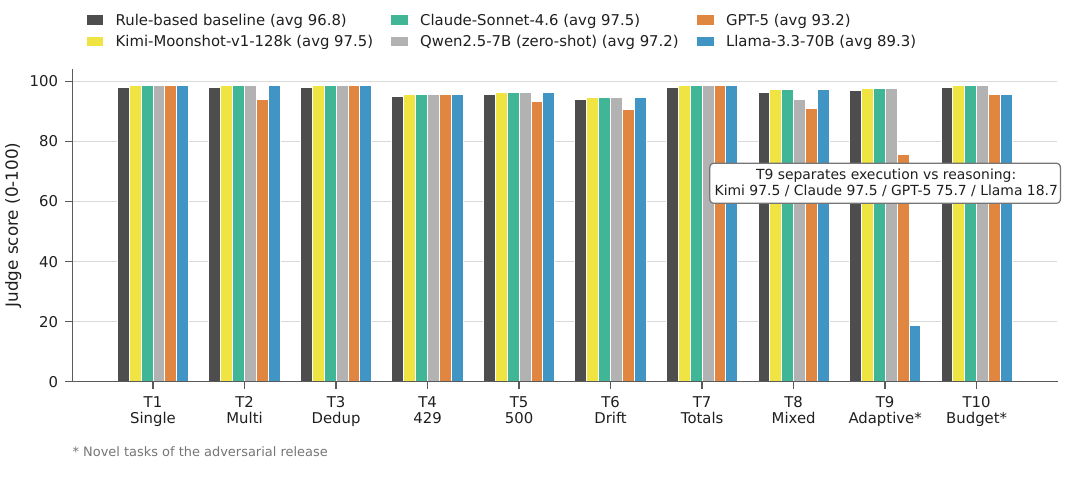}
\caption{Cross-model results on the original 10-task suite (T1--T10),
redrawn in this paper's figure style from the original release's per-run
score files (data unchanged; the original chart is preserved in the
repository). Scorer: original six-dimension judge, 0--100; scaffold: L0
production. Bars are per-task judge scores for six systems, including a
no-LLM rule-based baseline. Bars are near-identical across systems on
every task except the T9 adaptive adversary (and the rule-based script
sits inside the model pack): the flat-leaderboard signature of scoring
the scaffold rather than the model.}
\label{fig:legacy-benchmark}
\end{figure*}

The figures below are the artifacts of the original release
of the benchmark \citep{zhang2026comtradebench} and its
accompanying report. All benchmark numbers in them were produced by the
\emph{original six-dimension judge under the production scaffold}, the L0
configuration whose confound Section 3 (main text) diagnoses. We
reproduce them without alteration, for three reasons. First, completeness:
this is the full figure record, not a curated subset. Second, the flat
leaderboards visible in
Figures~\ref{fig:legacy-benchmark} and~\ref{fig:legacy-openenv1} are not a
historical curiosity but the confound in action: a no-LLM rule-based script
sits inside the model pack, and models of very different capability tie to
within a few judge points on T1--T8. Third, the GRPO figures
(Figures~\ref{fig:legacy-openenv2}--\ref{fig:legacy-training}) document a
training study run against the original L0 judge reward that is not discussed
in the main text; we preserve it because the observed
saturation-at-initialization and reward-variance collapse are themselves
evidence that the L0 reward carries little model-discriminating signal:
a policy can sit at the reward ceiling before a single gradient step.

\begin{figure*}[t]
\centering
\includegraphics[width=0.9\textwidth]{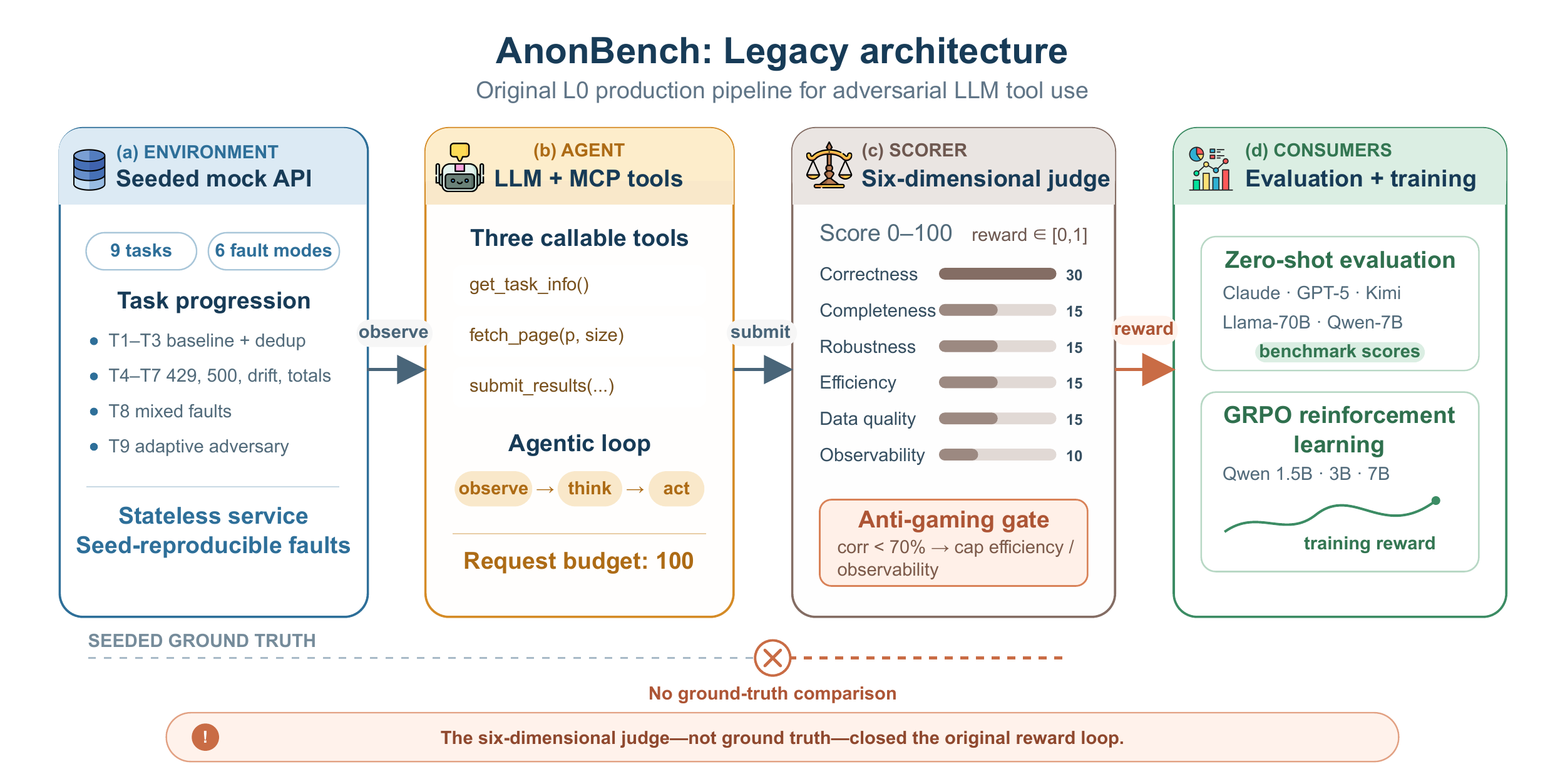}
\caption{Legacy architecture schematic of the original release (its footer
credit line was cropped in the review version; the uncropped original
ships in the artifact bundle): (a) the mock trade-statistics API serving
nine seeded tasks with synthetic fault modes, (b) the LLM agent loop over
three MCP tools under a request budget, (c) the six-dimension judge with
anti-gaming gates producing the 0--100 score used as reward, and (d) the
zero-shot evaluation and GRPO training consumers. Panel~(c) is the L0 scorer
diagnosed in the main text; the schematic documents that the judge, not
ground truth, closed the original reward loop.}
\label{fig:legacy-schematic}
\end{figure*}

\begin{figure*}[t]
\centering
\includegraphics[width=0.9\textwidth]{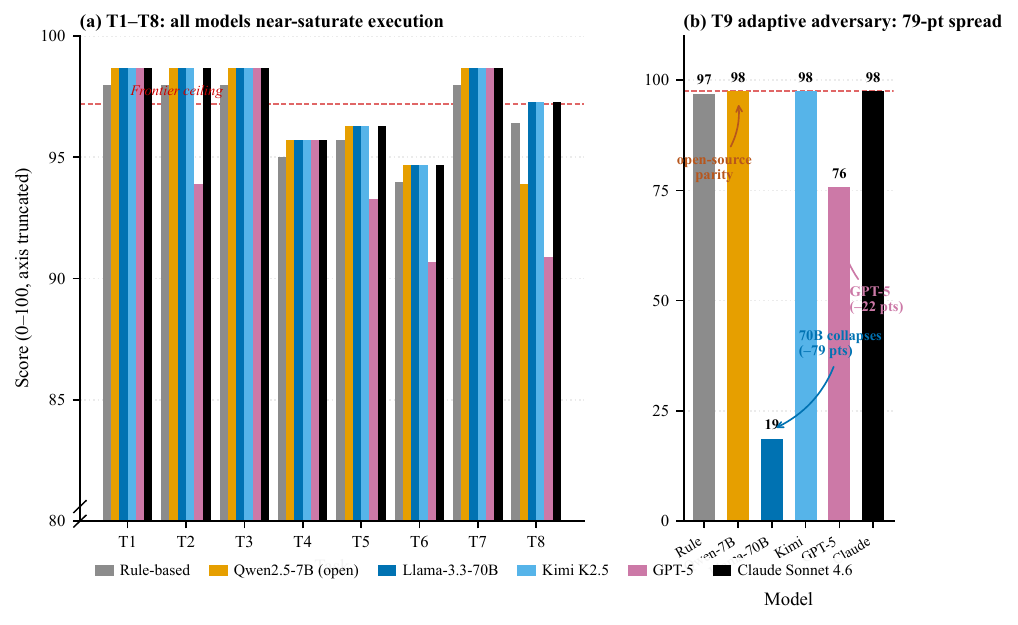}
\caption{Legacy Figure~1 of the original release (vector version): two-panel
benchmark chart, scored by the original six-dimension judge under the L0
production scaffold. Panel~(a): T1--T8 judge scores for six systems
(rule-based baseline included) on a truncated 80--100 axis, all crowding a
``frontier ceiling'' line, near-saturated execution. Panel~(b): the T9
adaptive-adversary task, the one task in the original suite that spreads the
systems, with an annotated 79-point spread. The original release read
panel~(b) as the interesting result; this paper's diagnosis is that
panel~(a) is the finding.}
\label{fig:legacy-openenv1}
\end{figure*}

\begin{figure*}[t]
\centering
\includegraphics[width=0.9\textwidth]{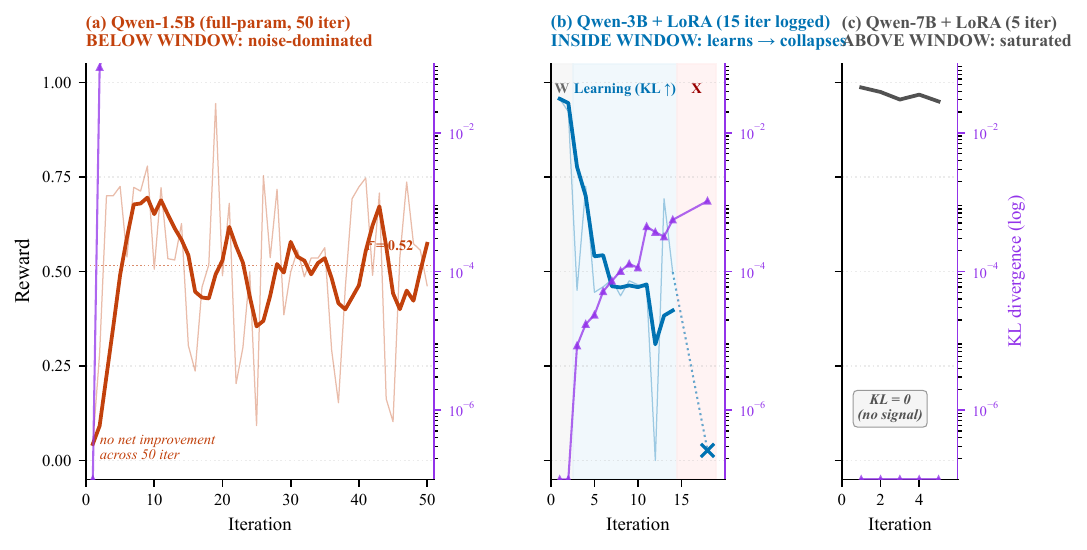}
\caption{Legacy Figure~2 of the original release (vector version): GRPO
training dynamics against the original L0 judge reward, for three model
scales. Panel~(a): Qwen-1.5B full-parameter, 50 iterations; reward
oscillates with no net trend (below the trainable window; noise-dominated).
Panel~(b): Qwen-3B$+$LoRA~\citep{hu2022lora}; reward and KL divergence move during a learning
phase, then the run collapses. Panel~(c): Qwen-7B$+$LoRA, 5 iterations;
reward starts at ceiling with $\mathrm{KL}=0$ (no gradient signal;
saturated). Not discussed in the main text; preserved because
saturation-at-initialization under the L0 reward is direct evidence that the
reward carries little model-discriminating signal.}
\label{fig:legacy-openenv2}
\end{figure*}

\begin{figure*}[t]
\centering
\includegraphics[width=0.9\textwidth]{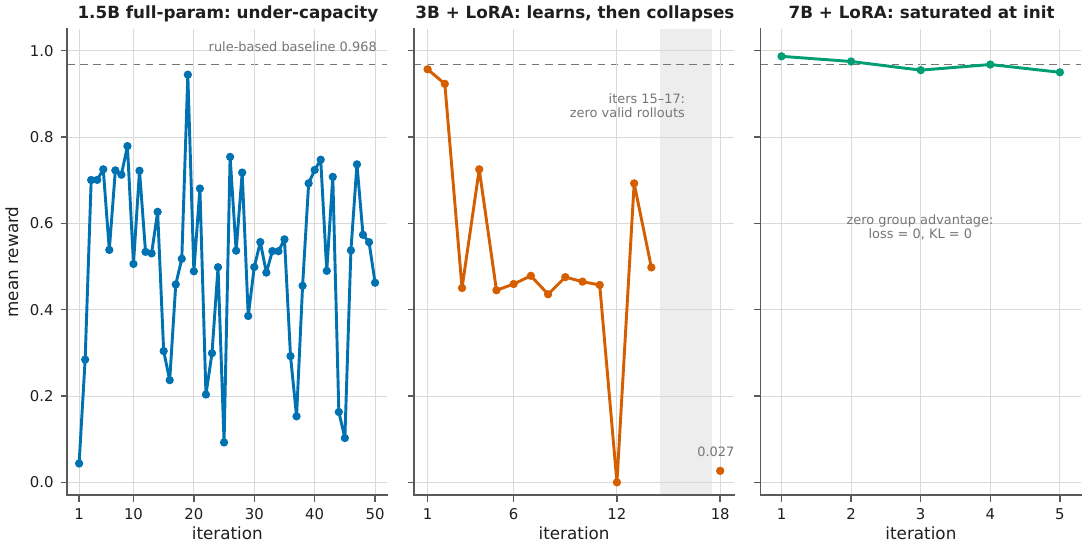}
\caption{GRPO operating envelope, regenerated from the raw per-iteration
records: mean reward per iteration under the original L0 judge reward for
the same three runs as Figure~\ref{fig:legacy-openenv2}, against the
rule-based baseline (0.968, dashed). The three empirically mapped failure
modes: under-capacity oscillation (1.5B full-param), learns-then-collapses
(3B$+$LoRA; iterations 15--17 produced zero valid rollouts, shaded, before
the final recorded point at 0.027), and saturated-at-init (7B$+$LoRA,
near-zero reward variance, hence zero GRPO advantage). The narrowness of the
usable band is a property of the reward, not of GRPO.}
\label{fig:legacy-envelope}
\end{figure*}

\begin{figure*}[t]
\centering
\includegraphics[width=0.9\textwidth]{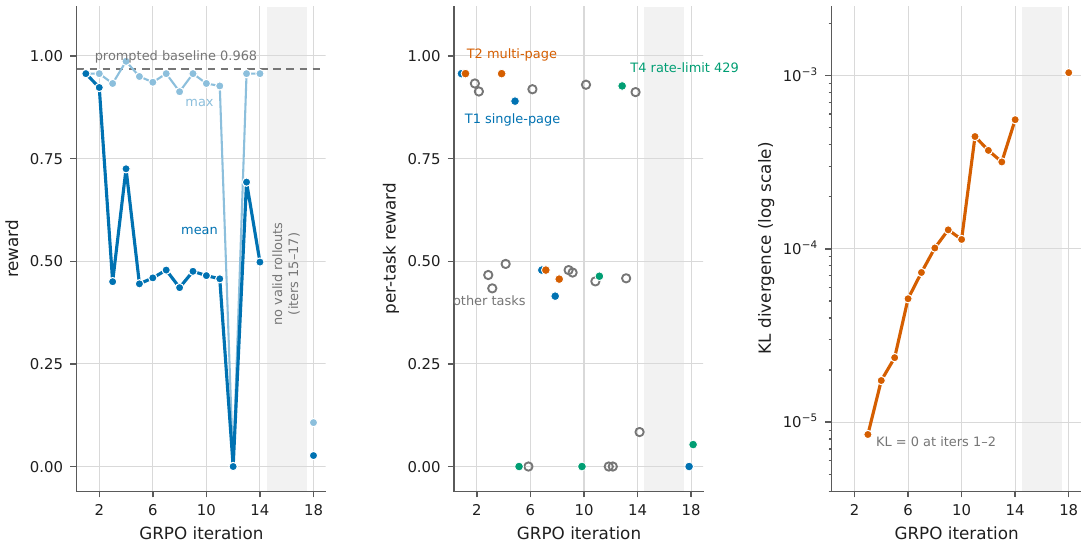}
\caption{GRPO training metrics for the 3B$+$LoRA run, regenerated from the
raw per-iteration records: mean and max reward per iteration (left),
per-task reward scatter (middle; the three most frequent tasks direct-labeled,
remaining tasks in gray), and KL divergence on a log scale (right; loss is
omitted; it oscillates around zero and carries no printable trend on a
single axis). Reward is the original six-dimension judge score (L0),
normalized. KL grows monotonically through the collapse (the adapter
keeps moving \emph{through} the degenerate region), and the per-task panel
shows the reward arriving as sparse task spikes rather than a graded
curriculum: context for Figures~\ref{fig:legacy-openenv2}
and~\ref{fig:legacy-envelope}.}
\label{fig:legacy-training}
\end{figure*}

\begin{figure}[ht]
\centering
\includegraphics[width=\columnwidth]{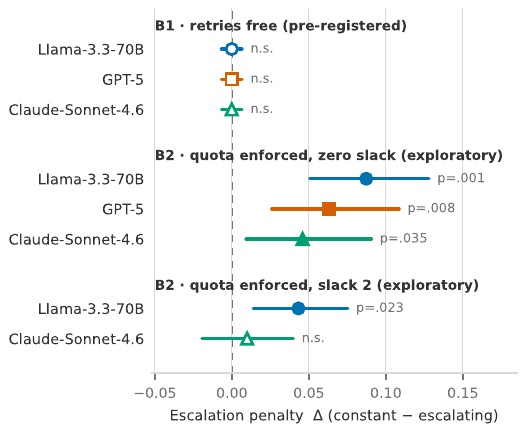}
\caption{Legacy copy of the paper's escalation-penalty forest plot (vector
version): paired difference $\Delta$ (constant $-$ escalating) in coverage
per model, with bootstrap intervals, for the pre-registered B1 null (retries
free; all n.s.), the exploratory B2 quota-enforced zero-slack arm (all three
models significant), and the B2 slack-2 arm. Metric: coverage under the L1
half-scaffold harness, not the judge. Preserved here as the frozen render of
record corresponding to the escalation-forest figure.}
\label{fig:legacy-ctb2}
\end{figure}

Where a figure was released in both vector and raster form
(Figures~\ref{fig:legacy-schematic}, \ref{fig:legacy-openenv1},
\ref{fig:legacy-openenv2}, \ref{fig:legacy-ctb2}) we
render the PDF; the PNG render of each ships alongside it in
\texttt{Figures/legacy/}. The benchmark-results chart
(Figure~\ref{fig:legacy-benchmark}) exists only as PNG; the two GRPO figures
(Figures~\ref{fig:legacy-envelope}, \ref{fig:legacy-training}) are
regenerated here in print form directly from the raw per-iteration training
records (generating scripts alongside the figures), with the original
release renders preserved in \texttt{Figures/legacy/}.

\end{document}